# Atmospheric electrification in the solar system

Karen L. Aplin,

Space Science and Technology Department, Rutherford Appleton Laboratory,

Chilton, Didcot, Oxon. OX11 0QX

E-mail: k.l.aplin@rl.ac.uk

**Abstract**

Atmospheric electrification is not a purely terrestrial phenomenon: all Solar System planetary atmospheres become slightly electrified by cosmic ray ionisation. There is evidence for lightning on Jupiter, Saturn, Uranus and Neptune, and it is possible on Mars, Venus and Titan. Controversy surrounds the role of atmospheric electricity in physical climate processes on Earth; here, a comparative approach is employed to review the role of electrification in the atmospheres of other planets and their moons. This paper reviews the theory, and, where available, measurements, of planetary atmospheric electricity which is taken to include ion production and ion-aerosol interactions. The conditions necessary for a planetary atmospheric electric circuit similar to Earth's, and the likelihood of meeting these conditions in other planetary atmospheres, are briefly discussed. Atmospheric electrification could be important throughout the solar system, particularly at the outer planets which receive little solar radiation, increasing the relative significance of electrical forces. Nucleation onto atmospheric ions has been predicted to affect the evolution and lifetime of haze layers on Titan, Neptune and Triton. Atmospheric electrical processes on Titan, before the arrival of the Huygens probe, are summarised. For planets closer to Earth, heating from solar radiation dominates atmospheric circulations. However, Mars may have a global circuit analogous to the terrestrial model, but based on electrical discharges from dust storms. There is an increasing need for direct measurements of planetary atmospheric electrification, in particular on Mars, to assess the risk for future unmanned and manned missions. Theoretical understanding could be increased by cross-disciplinary work to modify and update models and parameterisations initially developed for a specific atmosphere, to make them more broadly applicable to other planetary atmospheres.

## 1. Introduction and Scope

Comparative planetary science assumes that the science of Earth's environment can be used to understand other planetary environments; similarly, observations of other planets can be used to broaden understanding of the terrestrial environment. Several aspects motivate a modern comparative study of solar system atmospheric electricity. Firstly, there is evidence indicating that electrification could be a physical factor influencing Earth's climate. It appears likely that atmospheric charged particles could affect a planet's radiative balance through electrical influences on aerosol particle changes; both on Earth (Harrison and Carslaw, 2003) and elsewhere (Moses *et al.*, 1992). Secondly, studying other planetary atmospheres could contribute to understanding the origins of life, in which lightning has been implicated (Miller, 1953). Saturn's moon Titan is particularly interesting in this respect, as its atmosphere is thought to resemble the prebiotic Earth (Molina-Cuberos *et al.*, 2002). Thirdly, now that manned space missions to Mars appear probable (Bonnet and Swings, 2004), there needs to be an assessment of the potential electrostatic hazards facing future space missions.

The innermost planet, Mercury, and almost all the moons of the solar system will not be included in this paper, as they do not have atmospheres to become electrified (Lewis, 1997). Specifically, atmospheric electrification comprises lightning, which is caused by convection, and non-convective electrification. Non-convective electrification requires ionisation to produce electrically charged particles, which can originate from cosmic rays, radioisotope decay, or UV radiation. Cosmic rays are ubiquitous, so every planetary atmosphere can be expected to contain charged particles from ionisation, causing a slight atmospheric electrical conductivity. Through interactions with the ions and electrons produced by cosmic ray ionisation, aerosol particles add complexity to atmospheric charge exchange. The latitudinal distribution of atmospheric ionisation by cosmic rays is related to a planet's magnetic field, which modulates the incident cosmic radiation (Gringel *et al.*, 1986). In addition, the strength of the solar wind controls the deflection of cosmic rays away from a planet, and the effectiveness of this ionisation modulation over the solar cycle is related to the planet's distance from the Sun.

C.T.R. Wilson suggested that terrestrial atmospheric electrification was sustained by the existence of a global atmospheric electric circuit (Wilson, 1920), resulting from the electric current flow generated by disturbed weather and ionisation in the weakly conducting atmosphere between the surface and the ionosphere. (The ionosphere is a region of the upper atmosphere made electrically conductive by ionisation from absorption of solar UV radiation.) The terrestrial global atmospheric electric circuit concept is thought to be the best model available, despite some inconsistency with the limited number of observations (Dolezalek, 1972). The minimum parameters required for a global atmospheric electric circuit appear to be the existence of an atmosphere bounded by a conductive ionosphere and surface, with a charge generation mechanism. According to Harrison (2005), the charge generation in the alternating current (ac) part of a planet's global atmospheric circuit (mostly from lightning) can be identified by the presence of extremely low frequency (ELF) resonances in the surface-ionosphere waveguide: this also shows the presence of atmospheric charge generation and a conducting upper layer. Additionally, current flow is required to confirm the existence of a direct current (dc) global circuit. Both ELF signals and aspects of the ac global circuit are directly related to the occurrence of planetary lightning, which has been reviewed extensively (Rinnert, 1985; Desch *et al.*, 2002; Pechony and Price, 2004; Zarka *et al.*, 2004), and will not be covered in this paper beyond its association with convective and non-convective atmospheric electrification. Necessary conditions for a dc global atmospheric electric circuit to exist on a planet can be defined as:

1. Polar atmospheric molecules, to sustain charge.
2. Charge separation, usually generated by convection from meteorological processes, is required to form the dipole structure leading to electrostatic discharge. If there are no discharges, precipitation must carry charge to ground.
3. Evidence for conducting upper and lower layers.
4. Mobile charged particles, to provide current flow.

Discussion of planetary global circuits is still in its infancy, and only a few studies have considered the possibility of a global circuit on Mars (Fillingim, 1998; Farrell and Desch, 2001). Assuming that all planetary atmospheres have some charge separation due to ionisation, three aspects have been selected to characterise electrical systems in planetary atmospheres.

1. If convective or other processes cause sufficient charge separation for electrical discharges, meteorological changes will cause a global modulation of the planetary electrical system.
2. Aerosols, if present, are linked to local atmospheric electrification through ion-aerosol interactions.
3. If ions are implicated in the formation or removal of aerosol particles, there is a potential link between cosmic ray ionisation and the planet's atmospheric radiative balance.

This paper begins with a brief discussion of atmospheric electrification at Earth, to introduce the relevant physical processes. Table I gives an overview of solar system planetary atmospheres to be discussed in this paper, and their electrical systems, in terms of the three principal aspects listed above. Table I is organised moving outwards from the Sun from Venus to Pluto. Measurements and theoretical predictions of ionisation and other electrical processes in each planetary atmosphere are summarised, and the likelihood of a global atmospheric electric circuit on each planet discussed. Two moons with atmospheres and probable electrical activity have been included - a pre-Huygens survey of Saturn's moon Titan, and Neptune's largest moon Triton.

## 2. Fair-weather atmospheric electrification on Earth

### *2.1. The global electric circuit*

If the Earth's conductive ionosphere and surface are assumed to be spherical equipotentials, a potential difference of ~300kV exists between them, with the ionosphere the more positive. The ionospheric conductivity is $10^{-7}$-$10^{-3}$ $Sm^{-1}$, whereas the surface conductivity is variable depending on the water content, but at a minimum, dry rock conductivity is ~$10^{-6}$ $Sm^{-1}$ (Giraud and Petit, 1981; Lanzerotti and Gregori, 1986). The ionosphere-surface system has a finite capacitance with an *RC* time constant of a few minutes, indicating that there must be a constant current source to maintain the electric fields observed in the atmosphere (Harrison and Carslaw, 2003). Thunderstorms supply a net current by passing positive charge to the conductive upper atmosphere, and currents also flow from the surface up to the thundercloud (Rycroft *et al.*, 2000). Charge exchange from particle interactions (generally between ice crystals and soft hail) within a cumulonimbus (thunder) cloud is followed by charge separation from gravitational settling of negatively charged particles to form a dipole within the cloud. When the electric field exceeds the breakdown voltage of air, the cloud is discharged by lightning, either to the ground, within the cloud, or to other clouds (MacGorman and Rust, 1998). Electrical discharges directed upwards from the thundercloud to the ionosphere, called sprites and blue jets have recently been detected (Rycroft *et al.*, 2000). Electrostatic discharge and precipitation from electrified shower clouds comprise the disturbed weather part of the global circuit (Harrison, 2005). Figure 1 shows a diagram of the terrestrial global circuit, from Rycroft *et al.* (2000).

The total current ~2kA, averaged over the surface area of the Earth, gives a vertical current density, often referred to as the conduction current, $J_z$~2pAm$^{-2}$, carried by atmospheric ions, and an associated total global atmospheric resistance $R_T$~230Ω. Outside thunderstorm regions, the ionosphere-surface potential difference leads to a surface potential gradient PG~150Vm$^{-1}$. These conditions, away from charge separation processes, are often referred to as "fair weather". This paper follows the convention of defining PG as $+dV/dZ$, whereas the electric field $E = -dV/dZ$, so that in fair weather the PG is positive (Harrison and Carslaw, 2003). PG and $J_z$ are related to the air's electrical conductivity $\sigma$ (discussed in the next Section) by

$$PG = J_z / \sigma . \quad (1)$$

It should be noted that the typical fair weather conditions show considerable spatial variability; in particular, they are modulated by aerosol affecting the air conductivity, some of which arises from anthropogenic pollution (Harrison and Aplin, 2003).

The PG variation in clean air shows a characteristic Universal Time (UT) diurnal variation arising from the integrated thunderstorm activity on each continent. This variation was first detected on the cruises of the geophysical research ship, *Carnegie,* in the first half of the twentieth century (Roble and Tzur, 1986). The eponymous Carnegie variation has a broad peak in the afternoon and early evening UT. This is caused by the African thunderstorms dominating in the UT afternoon, and the early evening peak is related to the afternoon thunderstorm maximum in the Americas. Recent data analysis has revealed that the PG measured at clean air sites over land, where local aerosol concentrations are low, can also show the Carnegie variation (Harrison, 2004). The terrestrial Carnegie variation will be compared to a Martian diurnal electric field variation in Section 4.3 and Figure 6.

*2.2. Ionisation and atmospheric conductivity*

Terrestrial air conductivity arises principally from ionisation by cosmic rays. (Close to the continental surface, natural radioactive isotopes dominate.) The volumetric atmospheric ion production rate $q$ is latitudinally modulated by geomagnetic deflection; the magnetic field generated by Earth's core acts like a cosmic ray energy spectrometer, permitting only low energy cosmic rays to enter the atmosphere at high latitudes. Cosmic ray ionisation rates are therefore greater at higher latitudes, where a larger fraction of the cosmic ray spectrum contributes to ionisation (Gringel *et al.*, 1986). The solar cycle also modulates ionisation; at solar maximum the solar wind deflects more cosmic rays away from Earth, so the cosmic ray ionisation rate varies in antiphase with the 11-year solar cycle. Ionisation rate increases from the surface (~10 $cm^{-3}s^{-1}$, ~80% from natural radioisotopes emanating from the ground) up to the upper troposphere or lower stratosphere where the cosmic ray ionisation is strongest, 350 $cm^{-3}s^{-1}$ (Makino and Ogawa, 1985). The terrestrial $q$ profile is compared to that of Saturn's moon, Titan, in Figure 7. Typical surface air $\sigma$ values on Earth are $10^{-14}$-$10^{-15}$ $Sm^{-1}$, which is related to the mean ion mobility $\mu$ and the ion number concentration $n$ by

$$\sigma \approx ne\mu \,. \qquad (2)$$

In composition, terrestrial atmospheric ions typically consist of a central charged molecule stabilized by clustering with several ligands, commonly water or ammonia. Free electrons are unstable in Earth's atmosphere, and rapidly attach to electrophilic species to form negative ions. Atmospheric ions are lost by recombination and attachment to aerosol particles. The rate of change of bipolar ion number concentration $n_{\pm}$ can be represented as a balance between $q$, and two loss terms: self-recombination with coefficient $\alpha$ and attachment to a monodisperse aerosol population of number concentration $Z$ and an aerosol size-dependent attachment coefficient $\beta$. The steady-state ion balance equation can be written as

$$\frac{dn_{\pm}}{dt} = q - \alpha n_{+} n_{-} + \beta n_{\pm} Z \, . \qquad (3)$$

Once ions attach to aerosol particles, they are no longer electrically mobile enough to contribute to the atmospheric conduction current, or substantially to the air conductivity. However, there is evidence that charged aerosol can influence cloud microphysics by scavenging, the removal of aerosol by water droplets. Scavenging may be enhanced in certain regions by charging, which could affect the local cloud condensation nucleus (CCN) population or phase transitions (Tinsley *et al.*, 2000; 2001; Tripathi and Harrison, 2002) It has also been suggested that atmospheric ions can nucleate ultrafine aerosol particles, which would lead to an additional loss term in Equation (3) (Aplin and Harrison, 1999). There are two important atmospheric ion nucleation mechanisms. First, direct condensation onto ions is the process by which the ionising tracks of radioactive particles become visible in cloud chambers. In this paper, the direct nucleation mechanism is referred to as the “Wilson” mechanism, or ion-induced nucleation. This mechanism requires very much larger supersaturations than can be reached in the terrestrial atmosphere (Yu and Turco, 2001), but may be possible in the atmospheres of Venus (Section 3), Neptune (Section 7) and Triton (Section 8). Secondly, ion-mediated nucleation mechanisms have been postulated, in which charge indirectly enhances particle growth processes. Models predict ion-mediated nucleation in Earth’s atmosphere (Yu and Turco, 2001; Lovejoy *et al.*, 2004), and surface and free tropospheric observations provide evidence for the effect (Harrison and Aplin, 2001; Eichkorn *et al.*, 2002; Laakso *et al.*, 2004).

## 3. Venus

There has been much debate over the existence of lightning in Venus' dense, hot carbon dioxide atmosphere, covered with sulphuric acid clouds (Desch *et al.*, 2002). One of the few less controversial aspects is that the high atmospheric pressure, 9 MPa (90 bar) at the surface, increases the atmospheric breakdown voltage so much that it prevents cloud-to-ground lightning (Gurnett *et al.*, 2001). Recent radio observations from Cassini's flyby of Venus suggest that, if lightning does exist, its characteristics are unlike terrestrial cloud-to-ground and intracloud lightning, and may share radio emission characteristics with terrestrial sprites (Gurnett *et al.*, 2001).

### *3.1. Ionisation in the Venusian atmosphere*

Solar UV radiation cannot penetrate into the dense lower atmosphere of Venus, and cosmic rays are the principal source of ionisation at altitudes <60 km. Borucki *et al.* (1982) calculated cosmic ray ionisation rates for Venus based on a model verified in the terrestrial atmosphere, see Figure 2a (O'Brien, 1971; Borucki *et al.*, 1982). Surface ion production rates were ~0.01 ion pairs $cm^{-3}s^{-1}$, rising to ~0.5 $cm^{-3}s^{-1}$ at 30 km (1 MPa). Ionisation at 1000 hPa (50 km) was similar to terrestrial surface ionisation from cosmic rays. Venus has no magnetic field; thus, there will be no latitudinal effects on cosmic ray penetration into its atmosphere (Lewis, 1997). However, its proximity to the Sun causes a greater modulation in ionisation rates over the solar cycle than at Earth.

Borucki *et al.* (1982) argued that contributions to ionisation from radioactive rocks in the ground were likely to be insignificant outside a thin layer near the surface, on the basis that the upwards penetration of radioactive particles from the ground is related to atmospheric density. Scaling arguments were used to suggest that this contribution was confined to the lowest 100m, and cosmic rays were the only ionisation source considered in the model. However, radioactive uranium, thoron and potassium isotopes were detected at the Venusian surface by the Venera 8 mission $\gamma$-ray spectrometer in the 1970s (Vinogradov *et al.*, 1973). The proportions of radioactive elements in the rock were very similar to those found in the more radioactive terrestrial rocks like granite. The contribution of natural radioactivity to ionisation rates in the lowest 100m of the Venusian atmosphere can be estimated by assuming that the concentration of radioactive particles emitted from the ground is the same as on Earth. Composition differences between the Venusian and terrestrial atmospheres have a negligible effect on ionisation, as the energies needed to produce an ion pair in carbon dioxide and air are similar: 33.5eV and 35eV, respectively (Wilkinson, 1950; Borucki *et al.*, 1982). The terrestrial ionisation rate due to natural radioactivity, ~8 ion pairs $cm^{-3}s^{-1}$, can be scaled by the ratio of surface air pressures on Venus and Earth to give a figure for Venus ~0.01 ion pairs $cm^{-3}s^{-1}$. This would double the modelled ionisation rate in the 100m closest to the surface.

### *3.2. Ion-aerosol interactions*

Borucki *et al.* (1982) modelled ion clustering reactions and interactions with aerosol particles, in possibly the first theoretical study of extraterrestrial ion-aerosol interactions. Profiles of ion and electron concentrations, air conductivity and the charge density per unit volume (space charge) up to 80km were calculated, and key results are reproduced in Figure 2. In a dense atmosphere, primary ions and electrons rapidly form ion clusters. This process is similar to that on Earth, where water is an abundant clustering species, but on Venus the initial clusters are formed of carbon dioxide, which then quickly react with atmospheric trace gases in the warm conditions. Some of the final clusters formed are similar to those in the terrestrial atmosphere, particularly the hydrated cluster-ion, $H_3O^+(H_2O)_n$ (n = 3 or 4). Other common species are $H_3O^+(SO_2)$ and $H_3O(H_2O^+)(SO_2)$; the average positive ion mass is ~80 ± 40 amu. Less chemical information was available with which to model negative ion evolution, but sulphate species appear to dominate, with a mean negative ion mass ~150 ± 75 amu. Free electrons are also present above 60km, so there are three ion balance equations for Venus, with electrons and negative ions as two distinct species. The term representing positive ions is similar to the terrestrial case (Equation (3)), except that there are two recombination terms, one for negative ions and one for electrons; a similar approach can be used to model ion-electron physics in the terrestrial ionosphere (Ratcliffe, 1972). There is an additional loss term in the equation for negative ions representing electron detachment by collision, attachment and photo-induced mechanisms. The third ion balance equation, representing the rate of change of electron concentration, includes attachment terms for positive ions and aerosols, and an electron detachment term. A term to represent the attachment to neutral oxygen and sulphur dioxide molecules was included explicitly for negative ions and electrons.

Because of the ubiquity of cloud cover on Venus, attachment to aerosol particles is an important global ion loss mechanism. Venusian clouds exist from ~50-70km and are composed principally of sulphuric acid (James *et al.*, 1997). In the Borucki *et al.* (1982) model, a trimodal aerosol size distribution based on data from the Pioneer and Venera probes was assumed. Attachment coefficients were calculated using a simplified parameterisation for each mode, in each 1 km thick atmospheric layer. Ion and electron mobilities (Figure 2b) were obtained using the McDaniel and Mason (1973) method, whereas ion and electron concentrations (Figure 2c) were calculated by solving the three ion balance equations in the steady state. Equation (2) was used to compute conductivity, Figure 2d. Attachment significantly reduces ion and electron concentrations in the cloud layers. Borucki *et al.* (1982) predicted in-cloud charged particle concentrations of ~100 $cm^{-3}$, compared to ~$10^3$-$10^4$ $cm^{-3}$ if clouds were ignored. On Venus the lower conductivity at 50-70km is a global feature, Figure 2d. Generally, the atmospheric conductivity is approximately similar to that on Earth at similar pressures. Whilst there is uncertainty over the Venusian ion composition, positive ions appear to be more mobile than negative ones, in contrast to terrestrial ions where the chemistry favours clustering around positive ions, which reduces their mobility compared to negative ions.

### *3.3. Electrification and Venusian cloud formation*

In a detailed microphysics model, James *et al.* (1997) identified two cloud formation mechanisms. At the top of the cloud layer, particles are photochemically produced, but near the bottom, cloud formation is thought to result from heterogeneous nucleation. The atmosphere is supersaturated with respect to sulphuric acid from 40km upwards, where $H_2SO_4$ vapour can condense onto hydrated sulphuric acid particles, similar to terrestrial stratus generation by condensation of water vapour onto cloud condensation nuclei. As described in Section 3.2, Venusian atmospheric ions are likely to be sulphuric acid hydrates. The appropriate conditions may therefore exist for aerosols to form by ion-induced nucleation of sulphuric acid. The supersaturation required for ions to grow into ultrafine droplets by direct condensation can be determined using the Thompson equation (Mason, 1971). This equation describes the equilibrium saturation ratio needed for ion-induced nucleation to become energetically favourable. The equilibrium condition is defined at a saturation ratio *S,* where *r* is radius, $\rho$ fluid density, *M* the mass of the molecule, *q* charge, $\gamma_T$ the surface tension, $k_B$ Boltzmann's constant, *T* temperature, $r_o$ the initial radius (all in SI units), and $\varepsilon_r$ relative permittivity:

$$\ln S = \frac{M}{k_B T\rho}\left[\frac{2\gamma_T}{r} - \frac{q^2}{32\pi^2\varepsilon_0 r^4}\left(1-\frac{1}{\varepsilon_r}\right)\right]. \tag{4}$$

Equation (4) can be used to assess if condensation of gaseous $H_2SO_4$ onto ions is likely to occur in the lower cloud-forming regions at ~40km in the Venusian atmosphere. To calculate *S*, $H_2SO_4$ concentrations are required, and these were derived by Kolodner and Steffes (1998) from microwave absorption measurements made by the Magellan and Mariner spacecraft. Figure 3 shows an example $H_2SO_4$ profile. As Kolodner and Steffes (1998) also plotted the saturation concentration of $H_2SO_4$ ,$[H_2SO_4]_{sat,}$ the data can be used to determine saturation from the ratio of $[H_2SO_4]$ to $[H_2SO_4]_{sat}$.

$$S = \frac{[H_2SO_4]}{[H_2SO_4]_{sat}} \tag{5}$$

Temperatures at ~40km in Venus' atmosphere are ~400K, and were determined as a function of altitude using a linear regression from the model atmosphere in Borucki *et al.* (1982). Surface tension, density and dielectric constant data for ~100% sulphuric acid were not available at 400K. Instead, the surface tension at 300K, 0.0725 $Nm^{-1}$ was used (Myhre *et al.*, 1998). Densities of $H_2SO_4$ at 300K were quoted by Myhre *et al.* (1998) for mass fractions of water from 0.123-0.765. A fourth-order polynomial fit to the density variation with mass fraction at 300K was used to estimate the density of 100% sulphuric acid at 300K to be 1000 $kgm^{-3}$. There was insufficient information to extrapolate the relative dielectric constant of $H_2SO_4$ to high temperatures, so the measured value of 101 at 298K was assumed (Gillespie and Cole, 1956). Equation (4) was then used to compute the saturation ratio needed for nucleation onto ions with 1, 2 and 5 electronic charges at a temperature $T$, at the different Venusian locations for which $H_2SO_4$ profiles were retrieved. The results are summarised in Table II.

As discussed in Section 2.2, "Wilson" nucleation is impossible in the terrestrial atmosphere because water supersaturations rarely exceed a few percent, and $S$=4 is required for condensation onto ions. Sulphuric acid has a lower vapour pressure and a higher permittivity (polarisability) than water, so the saturations required for condensation onto singly charged Venusian ions are lower than on Earth. Despite this, $S$ is never high enough for condensation onto particles with one electronic charge. From Table II, the lowest supersaturations[1] are needed for nucleation on Venus at 88°S, at 42.5-43.5 km and 46.8-47.5 km (Figure 3). The maximum SS, 85%, occurs at 47.2km. As this is relatively cold (370K), and the SS needed for Wilson nucleation is inversely related to temperature (Equation (4), the most probable location for ion-induced nucleation is ~43 km. In this region there is lower SS, 25-30%, but, with the higher temperatures (~395K), the SS needed for condensation onto charged particles is also lower. This is illustrated in Figure 4a which shows the solution of Equation (4) at 395K. Doubly charged particles of radius 1nm can nucleate at SS=7%, and particles with 5 charges of radius 2nm can nucleate at SS=1-2%. These SS appear to be relatively easily attained at 88°S, based on Figure 3.

[1] Supersaturation occurs when $S > 1$. $S = 1.07$ is equivalent to a supersaturation (*SS*) of 7%.

The possibility of doubly charged particles existing in this region can be estimated from the aerosol charge distribution arising from ion-aerosol attachment processes. The steady state charge distribution of a monodisperse aerosol population, represented as the ratio of the number of particles with charge *j*, $N_j$, to the number of neutral particles, $N_0$, can be given by

$$\frac{N_j}{N_0} = x^j \frac{8\pi\varepsilon_0 akT}{je^2} \sinh\left[\frac{je^2}{8\pi\varepsilon_0 akT}\right] \exp\left[\frac{-j^2e^2}{8\pi\varepsilon_0 akT}\right] \qquad (6)$$

where *j* is the electronic charge, *a* the mean aerosol radius, and *x* is the ion asymmetry factor, $\frac{n_+\mu_+}{n_-\mu_-}$ (Clement and Harrison, 1992). The charge distribution can be calculated for T=395K and typical ion concentrations and mobilities at 40km on Venus. The mobility of positive ions was assumed to be 7% higher than negative ions as quoted in Borucki *et al.* (1982). Venus' atmosphere is too dense for free electrons to exist below 60km, and their contribution to the ion balance can be ignored. Positive and negative ions were assumed to exist in equal concentrations (Figure 2c) so *x*=1.07 from the difference in mobility. The aerosol diameter at 40km was assumed to be 0.25μm following Borucki *et al.* (1982). Substituting these values of *a, T* and *x* in Equation (6) indicates that 84% of aerosols in this region carry some charge, with the most common single charge state being +1. The mean charge is +0.2 (equation 19 in Harrison and Carslaw, 2003), because of the higher mobility of positive ions, but, across the charge distribution, Figure 4b, $j \geq 2$ is relatively common. According to the estimated charge distribution, 54% of Venusian aerosols carry enough charge for ion-induced nucleation at supersaturations of 7%, and 7% of the aerosol population carries enough charge for nucleation at 1-2% supersaturation. These estimates suggest that it may be possible for gaseous sulphuric acid to condense onto charged aerosol particles below the cloud layer in Venus' atmosphere.

### *3.4. Is there a global electric circuit on Venus?*

At the moment, the existence of a Venusian global electric circuit seems unlikely due to the lack of conclusive evidence for current flow through precipitation or electrostatic discharges. Pechony and Price (2004) predict that any ELF resonance resulting from lightning on Venus would result in a well-defined peak at ~9Hz. This may offer another approach to searching for Venusian lightning. If the existence of lightning is later proven, then there is a good basis for the existence of a global circuit if the Venusian surface is more electrically conductive than its atmosphere. The composition of the Venusian surface is not yet well known because the cloudy, dense atmosphere makes remote sensing of the surface very difficult. Most of the data is based on results from the three landers, Venera 13, Venera 14 and Vega 2, which all landed on a similar, basaltic, region of the planet (Lewis, 1997). Radar data transmitted by the Magellan spacecraft in orbit around Venus suggested that the electrical conductivity of the Maxwell Montes highland region was ~$10^{-13}$ $Sm^{-1}$ (Pettengill *et al.*, 1996). As the average air conductivity on Venus is ~$10^{-14}$ $Sm^{-1}$ then, if this value is representative, the ground could be more conductive than the air. The existence of heavy metal frosts on the higher surfaces has been postulated, such as tellurium and, recently, lead and bismuth compounds (Schaefer and Fegley, 2004). Even though lead and bismuth are relatively poor conductors of electricity, the ground conductivity should be increased by their presence, which could increase the possibility of a global circuit. Further data are required to determine whether lightning occurs on Venus, and what form it takes.

### *3.5. Future Missions*

Venus Express, a European Space Agency orbiter, is the next mission to Venus, with launch planned for November 2005 (ESA, 2003). Its aim is to study the atmosphere, map surface temperatures and measure the interaction of the atmosphere with the solar wind, with instrumentation principally based on the successful Mars Express mission. Whilst new atmospheric profile and composition data will undoubtedly be of use for improving models of Venusian ion-aerosol physics, the spacecraft carries little specific atmospheric electrical instrumentation. Its magnetometer instrument may be able to contribute to lightning detection experiments, as will its cameras. Venus Express will also carry a Planetary Fourier Spectrometer which will scan the atmosphere between 0.9-45μm, constraining details of known atmospheric constituents and perhaps identifying new species. It will also characterise aerosol composition and size distributions from their optical properties. The aerosol data could provide a more detailed size distribution to be used in ion-aerosol models like that of Borucki *et al.* (1982), and the ion-induced nucleation estimates presented in Section 3.3.

Another European mission, Lavoisier, was unsuccessfully proposed to ESA in 2000 (Chassefière *et al.*, 2002). This was based on Huygens technology (to be discussed in Section 6) and, like Huygens, one of its objectives was to search for atmospheric electrical activity. Its descent probe and three balloon flotillas were to include a relaxation probe to measure atmospheric conductivity and acoustic sensors for lightning detection. An alpha particle detector would measure the surface ionisation rate from natural radionuclides to detect crust outgassing and volcanism. It is unfortunate for the study of Venusian atmospheric electrification that Lavoisier was not selected for further development.

## 4. Mars

The closeness of Mars to Earth has meant that meteorological and geophysical data have been acquired from several space missions. Unfortunately none has included electrical instrumentation, and the only experimental evidence for electrical activity on Mars is the electrostatic adhesion of dust to the wheels of the Mars Pathfinder and Sojourner rovers (Farrell *et al.*, 1999; Ferguson *et al.*, 1999; Berthelier *et al.*, 2000). However, based on deductions from the terrestrial analogues in desert sandstorms, Mars is expected to have substantial atmospheric electrification.

Mars is unique in the solar system because of the significance of dust in its climate. The fine dust grains on the surface can be suspended by wind to form dust storms which can sometimes cover large areas of the planet. Convective vortices, analogous in structure and size to terrestrial dust devils, also play an important part in lofting dust into the atmosphere. One consequence of the dust's low thermal inertia (Lewis, 1997) is that the surface responds rapidly to solar radiation, leading to large diurnal variations in temperature. Atmospheric dust loading could have important radiative effects on Mars; meteorological models developed from terrestrial general circulation models are being used to investigate the role of dust in Martian global climate (Newman *et al.*, 2002). Recent findings suggest that temperatures in the Martian northern hemisphere spring and summer could be controlled by convectively lofted dust (Basu and Richardson, 2004).

### *4.1. Ionisation and atmospheric conductivity*

Unlike Earth, Mars does not have a substantial global magnetic field with which to deflect cosmic rays away from the planet. The Martian equatorial magnetic field is 0.5nT compared to 30-60μT (equator and poles, respectively) for Earth (Tribble, 1995; Connerney *et al.*, 2001), arising from magnetization of the martian crust. The Mars Global Surveyor instrument carried a magnetometer, from which global maps of the Martian magnetic field were produced (Connerney *et al.*, 2001). It indicated that most magnetization was concentrated in the southern hemisphere with maximum radial fields reaching 200nT in the Terra Cimmeria/Terra Sirenum region, centred on ~60°S, 180°. The Martian cosmic ray flux is likely to be higher than Earth's in its less dense atmosphere, with no latitudinal modulation, but the regional magnetic field might modulate ionisation rates to be slightly higher in the northern hemisphere. In the daytime, solar UV photons can penetrate to the surface, as, unlike Earth, there is no atmospheric ozone layer to absorb them (Fillingim, 1998). They ionise the surface to leave it slightly positively charged, and the emitted photoelectrons contribute to atmospheric conductivity and ionise the Martian air (Grard, 1995). Typical surface daytime free electron densities are 1-100 $cm^{-3}$, dropping to <1 $cm^{-3}$ at night. The free electrons and ions formed by electron impact ionisation cause high air conductivity at the surface, ~$10^{-11}Sm^{-1}$, comparable to terrestrial stratospheric conductivity (Berthelier *et al.*, 2000). The positive conductivity is small because of the dominance of atmospheric $CO_2$, which forms negative ions, and highly mobile electrons. Nitrogen is the next most abundant atmospheric species (Table I), and forms positive ions, but in a carbon dioxide atmosphere, $N_2^+$ reacts rapidly with $CO_2$ to form $CO_2^+$ and $N_2$ (Fehsenfeld *et al.*, 1966). At a relative abundance of only 2.7% the contribution of $N_2^+$ to Martian conductivity would be negligible even if the ions formed are chemically stable. One interesting consequence of photoionisation is that there is likely to be a large diurnal variation in atmospheric conductivity, to be discussed in Section 4.4.

### *4.2. Electrical discharges*

Electrical discharges have not yet been detected on Mars, but it is widely supposed that they occur. Up to $10^6$ elementary charges $cm^{-3}$ (0.1 $pCcm^{-3}$) and electric fields of $kVm^{-1}$, generated by triboelectric charging, have been measured in terrestrial dust devils (Farrell and Desch, 2001). Calculations and laboratory measurements indicate that high electric fields are also expected in Martian dust devils (e.g. Krauss *et al.*, 2003). Farrell *et al.* (1999) suggest that electric fields generated in a dust devil are likely to be limited by the breakdown voltage of the low-pressure Martian atmosphere, ~$10kVm^{-1}$, rather than the maximum charge sustained by an individual dust grain. They predicted that a dust devil 5km across could sustain ≤200 elementary charges $cm^{-3}$ (32 $aCcm^{-3}$) before breakdown. Two breakdown mechanisms were suggested, firstly a corona or glow discharge due to local breakdown of the air at the edge of the dust devil, which self-limits the maximum field attained, and secondly a spark discharge similar to terrestrial volcanic lightning (Farrell *et al.*, 1999; Farrell and Desch, 2001).

*4.3. Martian global circuit*

Farrell and Desch (2001) have discussed the possibility of a Martian atmospheric electrical circuit with electrical discharges in dust storms as current generators, similar to thunderstorms on Earth. As defined in Section 1, a global circuit requires the existence of a conductive ionosphere and surface, charge separation and a conductive atmosphere. The Martian ionosphere extends upwards from ~15km, much lower than on Earth because of the presence of free electrons in the lower atmosphere (Berthelier *et al.*, 2000). The lack of water in the Martian regolith makes the surface a relatively poor conductor. In their model Farrell and Desch (2001) assumed that the ground conductivity >$10^{-9}$ $Sm^{-1}$. Based on the assumption that Martian charge separation processes in dust clouds are not as efficient as theoretical considerations suggest, as for terrestrial dust devils, they estimated the "fair weather" electric field and conduction current density generated by dust storms. In the Martian context, "fair weather" refers to areas not covered by a dust storm. In the storm season (northern hemisphere winter) one very large regional (500x500x20 km) storm is expected, with several medium-sized (50x50x15 km) storms. The resistance in the air column above the dust storm was estimated from electron density models, and the voltage in the dust cloud was assumed to be $1kVm^{-1}$ multiplied by the cloud height. Ohm's Law was then used to calculate the current contributed to the global circuit: ~2kA for the regional storm and ~500A in total from the moderate storms. This current is averaged over the fair weather area to give an electric field $E$ = 475 $Vm^{-1}$ and a conduction current of $J_z$ = 1.3 $nAm^{-2}$. Outside the storm season, dust devils at the surface are common, and the only source of charge for the global circuit. If each is estimated to contribute 1A then this gives $E$ = 0.14 $Vm^{-1}$ and $J_z$ = 0.4 $pAm^{-2}$. The daytime fair weather electric field may be enhanced by 0.01-0.1 $Vm^{-1}$ by the emission of photoelectrons from the surface (Grard, 1995). The atmospheric conductivity calculated using these estimates from Equation (1), ~$10^{-12}$ $Sm^{-1}$, is inconsistent with the estimate of ~$10^{-11}$ $Sm^{-1}$ in Grard (1995) and Berthelier *et al.* (2000); this will be discussed in Section 4.4. A schematic diagram of the model of Farrell and Desch (2001) is shown in Figure 6a. Note that the current flow and electric fields are oppositely directed to those in the terrestrial global circuit, Figure 1.

In the absence of direct measurements, there are substantial uncertainties in the Farrell and Desch (2001) model. Discharge via local corona currents would not contribute to the global circuit, unlike sparks to the ground or upper atmosphere. The likelihood of local corona discharge is dependent on the efficiency of the tribocharging process; if it is more efficient than assumed by Farrell and Desch (2001), coupling to the global circuit will drop, reducing $E$ and $J_z$. The conductivity of the Martian surface is also poorly understood; values in the literature from $10^{-7}$ to $10^{-11}$ $Sm^{-1}$ were discussed by Berthelier *et al.* (2000) who believed the best estimate was $10^{-10}$ to $10^{-12}$ $Sm^{-1}$. The ratio of ground to air conductivity $\sigma_g/\sigma_a$ is ~$10^4$-0.1: according to the constraints defined in Section 1, $\sigma_g/\sigma_a$ must be ≥1 for a global circuit (on Earth $\sigma_g/\sigma_a$~$10^9$). If the atmospheric conductivity is ~$10^{-11}$-$10^{-12}$ $Sm^{-1}$ then a Martian global circuit seems less likely; however, both atmospheric and surface conductivity are poorly constrained. It appears that the limits of knowledge from the small amount of Martian atmospheric electricity data are being approached.

### *4.4. Variability in Martian atmospheric electricity*

Martian fair weather atmospheric electricity is likely to be much more variable than on Earth, for reasons first summarized by Fillingim (1998). Diurnal variations are substantial, and photoelectrons almost certainly dominate the Martian daytime atmospheric conductivity (Equation (2)) because of their high number concentration and electrical mobility, given as 200 $m^2V^{-1}s^{-1}$ in $CO_2$ at 1.33hPa (Hasegawa *et al.*, 1998). Martian atmospheric $CO_2^-$ ions are orders of magnitude less mobile than the electron. Even if electrons and $CO_2^-$ exist in equal concentrations, electrons control >99.5% of the Martian surface air conductivity. At night there is no photoelectron emission, and charged particles are only formed by cosmic ray ionisation, causing a large diurnal air conductivity variation. The value of $10^{-11}$ $Sm^{-1}$ estimated by Grard (1995) is based on photoelectrons, whereas the air conductivity calculated from Equation (1) of $10^{-12}$ $Sm^{-1}$ is an average value, including the much lower nocturnal conductivity. The air density on Mars varies by 20% over the year (Lewis, 1997), leading to seasonal variations in cosmic ray penetration into the atmosphere. Cosmic ray ionisation rates in the Martian lower atmosphere are not well known, but the daytime air conductivity near the surface may not be significantly affected by the seasonal pressure variation because of the dominance of photoelectrons.

Diurnal and spatial variations in the storms contributing charge to the global circuit modulate the daily cycle of fair weather electric field, described in Section 6. If a global circuit exists on Mars then the planet may have its own "Carnegie" diurnal electric field variation. This depends on both the location of the dust storms contributing to the global circuit and their diurnal variation. If these regions and variations can be identified, then it is possible to predict a Martian "Carnegie curve" using data on the diurnal variation of dust devils, and the distribution of large dust storms. Ringrose *et al.* (2003) have analysed the Viking 2 lander meteorological data from 1976, and identified convective vortices passing near the lander. From this they present a diurnal variation in dust devil activity peaking at about 1300 local time, Figure 5a. This is similar to the pattern of terrestrial convective vortex generation, with an unexpected morning peak. The starting locations of large Martian dust storms from 1894-1990 have been plotted in Newman *et al.* (2002); these data are presented in simplified form in Figure 5b, showing the number of storms in 30º longitudinal bins. The diurnal variation in relative intensity of atmospheric electric field at 0º longitude has been computed based on the dust storm diurnal variation at different Martian local times, and this Martian "Carnegie curve" is plotted in Figure 6b. It shows a bimodal pattern with a nocturnal minimum. This is a consequence of the lack of dust devils at night, and the scarcity of storms at longitudes with a 10-12 hour time difference from 0º. The morning peak results from the dust storm activity around Hellas Planitia at 60-130º E, and the afternoon peak from the strongest single dust storm area, in the Xanthe Terra/Marineris region (Greeley and Batson, 2001). The shape of the curve was relatively insensitive to the magnitude of the morning peak in dust devil activity, which may be of instrumental origin (Ringrose *et al.,* 2003). Figure 6b is an estimate based on the limited data available, and assumes that all Martian dust storms, irrespective of size, have the same diurnal variation as dust devils, and the spatial variation of the starting location of large storms. All storms are also assumed to contribute equally to the global circuit. The Martian Carnegie curve is expected to have a substantial seasonal variation, as indicated by Farrell and Desch (2001), because of seasonal changes in dust storm frequency.

Mars has variable topography: its highest peak, the biggest mountain in the Solar System, Olympus Mons, is 25 km high (Lewis, 1997). At Olympus Mons and the nearby Tharsis range, the distance from surface to ionosphere is substantially reduced, which should increase the local conduction current (Fillingim, 1998). The variations of magnetic field over Mars are likely to have complex consequences near the more magnetic regions. Fillingim (1998) pointed out that atmospheric thermal tides may drag ionospheric plasma across magnetic field lines, inducing electric fields and currents. Local cosmic ray penetration could be modulated, though this appears unlikely as even in the most magnetic regions the Martian magnetic field is two orders of magnitude lower than Earth's magnetic field (Molina-Cuberos *et al.*, 2001a).

### *4.5. Future mission plans*

Mars is arguably the most explored body in the solar system, having received several visits since Mars 2, the 1971 Russian orbiter which was the first successful mission to the red planet (Lewis, 1997). The absence of electrical instrumentation on any of the missions severely limits our understanding of the Martian electrical environment. The lack of quantitative measurements of Martian atmospheric electrification could be a constraint on future missions, especially as Mars will almost certainly be the first extraterrestrial planet to be visited by humans, and all potential hazards must be quantified before this, to minimise risk. Electrostatic measurements could also be used to improve the performance of future unmanned missions to Mars, as robotic explorers are also threatened by electrostatic discharges. Pechony and Price (2004) predicted that any ELF resonances resulting from Martian electrostatic discharges would not be well-defined, so the most promising approach for detecting atmospheric electrical activity appears to be through *in situ* instrumentation. Calle *et al.* (2004) describe electrostatic sensors embedded in the wheels of rovers, measuring the charge induced on a metal electrode by soil or dust particles. The variation in charging with the soil material could provide mineralogical as well as electrical information. Other Martian atmospheric electrical instrumentation has been proposed, such as a sensor to detect both atmospheric electric fields and conductivity, and an extremely low frequency AC field detector to respond to electrical discharges (Berthelier *et al.*, 2000). A prototype spectrometer to detect the charge and size distribution of atmospheric dust has also been developed (Fuerstenau and Wilson, 2004). There are still no definite plans for electrostatic instrumentation on any future Martian missions.

## 5. Jupiter and Saturn

Jupiter and Saturn both resemble the Sun more than the Earth. Their huge atmospheres are dominated by hydrogen and helium, in which the pressure and temperature increase with depth. At pressures > 40GPa, hydrogen starts to dissociate, and becomes "metallic" and electrically conducting at pressures > 300GPa. It is probable that Jupiter's "surface" is a high-pressure liquid hydrogen "ocean" which becomes semi-conducting and then conducting as the pressure increases, with a small (Earth-sized) rocky core (Nellis, 2000; Stevenson, 2003). The metallic hydrogen acts as a dynamo generating the substantial Jovian planetary magnetic field, which is the highest of any body in the solar system, see Table III. Despite these apparent differences from the terrestrial planets discussed above, Jupiter and Saturn have active weather systems driven by convection. Unlike the inner planets, where weather systems are driven by insolation, convection in the outer planets arises principally from internal heat generation (Atreya, 1986).

Jupiter in particular has been the subject of many years of study, as its size makes it relatively easy to observe surface features from Earth. For example, Robert Hooke reported observing the motion of bands and spots on Jupiter in 1666 (Hooke, 1666). Both Saturn and Jupiter have active convection and cloud layers composed of ammonia, ammonium hydrosulphide and water, resulting in their characteristic colours and banded structure (Lewis 1997; Desch *et al.*, 2002). Clouds and convection suggest lightning activity, which was predicted on Jupiter in the 1970s, and might also be expected on Saturn, which has a similar atmosphere (Desch *et al.*, 2002). Jovian lightning was optically detected by Voyager and Galileo (Lanzerotti *et al.*, 1983; Little *et al.*, 1999), and Voyager observed radio signals from lightning on Saturn (Kaiser *et al.*, 1983). Jovian lightning probably originates from water clouds deep within the atmosphere, and is generated by similar mechanisms to terrestrial lightning (Lewis, 1997; Yair *et al.*, 1998). From observations of the rotation period of the signals, Kaiser *et al.* (1983) associated the "Saturn electrostatic discharges" (SEDs) with equatorial storms, and this connection was confirmed by recent Cassini observations. The Radio and Plasma Wave Science (RPWS) instrument observed SEDs (Gurnett *et al.*, 2005) which were spatially and temporally correlated with southern mid-latitude storm systems seen by the Imaging Science Subsystem (Porco *et al.*, 2005a). Whistler mode emissions were also seen by the RPWS package and are thought to emanate from Saturn's rings (Gurnett *et al.*, 2005).

### *5.1. Cosmic ray ionisation*

Jupiter's strong magnetic field screens out all except the most energetic cosmic rays. High-energy particles of >100MeV may influence chemical reactions in the lower atmosphere where solar UV radiation cannot penetrate (Lewis, 1997). This motivated the study of Capone *et al.* (1979) who modified an earlier cosmic ray atmospheric ionisation model (Capone *et al.*, 1976) to include the effect of muons, which are relatively more important at atmospheric densities > 750gcm$^{-2}$ (~750hPa). Cosmic ray ionisation was predicted to occur at altitudes down to 40km below the visible clouds, after which the dense atmosphere absorbs all the cosmic rays. Terminal positive ions were $NH_4^+$ and $C_nH_m^+$ cluster-ions; $H^-$ ions are formed, but react so quickly that their concentration is effectively zero. Modelling the ion chemistry indicated that cosmic ray ionisation could ultimately enhance synthesis of molecules like $C_3H_8$ (propane) and $CH_3NH_2$ (methylamine/amino methane), but solar UV ionisation would dominate above the tropopause (Lewis, 1997; Capone *et al.*, 1979). Because of the large magnetic field, the equilibrium electron density is relatively sensitive to geomagnetic latitude; it is a factor of 4.5 higher at the poles than at the equator, suggesting that any Jovian cosmic ray chemistry could show a strong latitudinal variation. A similar model applied to Saturn's atmosphere found that $C_2H_9^+$ dominated in the lower atmosphere, but no chemical predictions were made (Capone *et al.*, 1977), possibly because Saturn's atmosphere has a higher relative concentration of hydrogen and a lower abundance of reactive trace species than Jupiter (Lewis, 1997).

### *5.2. Global atmospheric electric circuit*

Whilst both lightning and cosmic ray ionisation exist in the atmospheres of Jupiter and Saturn, there are two reasons why the structure of a gas giant planet prevents a global electric circuit. First, cosmic ray ionisation decreases as atmospheric pressure increases, and a region of the deep atmosphere where ionising species cannot penetrate is predicted (Capone *et al.*, 1979). Charged species could only enter this region by transport downwards, which is unlikely due to the internal heat source driving convection, and suggests the deep Jovian atmosphere may be electrically neutral. Another condition for a global circuit is that a planetary surface needs to be more conductive than its atmosphere (Section 2.1). Hydrogen is compressed into an insulating liquid before it starts to dissociate and become electrically conductive, so the Jovian "surface" is probably electrically insulating (Stevenson, 2003). Sentman (1990) predicted that lightning would cause resonances at 1-2 Hz in the cavity between Jupiter's surface and ionosphere. However, more recent laboratory work has improved understanding of the behaviour of hydrogen at high pressure (Weir *et al.*, 1996), and Sentman's (1990) assumptions about the conductivity of the Jovian interior may now be outdated. It is concluded that not all the necessary conditions for a global circuit, a conductive atmosphere, and a conductive surface relative to the atmosphere, appear to be fulfilled for Jupiter and Saturn.

## 6. Titan

The atmosphere of Saturn's largest moon, Titan, has long been of interest to astronomers and planetary scientists, because it is thought to resemble conditions on Earth several billion years ago. Very little was known about it until the Voyager 1 flyby in 1980. Its composition is principally nitrogen with 6% methane. The temperature range is such that methane exists in three phases, much like water on Earth. This has led to suggestions that there is a "hydrological" methane cycle on Titan with methane clouds and rain, although the methane raindrops may evaporate before they reach the ground (Tokano *et al.*, 2001a). There are numerous trace organic species in Titan's atmosphere suggesting that electrical discharges may have contributed to their formation (e.g. Capone *et al.*, 1980; Lewis, 1997; Desch *et al.*, 2002; Lorenz, 2002). Understanding of this moon's meteorology has been hindered by the dense stratospheric haze which obscures the surface and almost totally prevents remote sensing of the lower atmosphere. Adaptive optics technology has improved observations from Earth and provided some information on surface structure (Zarnecki *et al.*, 2002; de Pater, 2004). However, the surface properties remain poorly understood, and there has been intense speculation about the possibility of methane lakes, oceans, and even oily sludge (Lorenz, 2003). Titan has a relatively conductive ionosphere, compared to its lower atmosphere and, whilst lightning has not been observed, electrification within methane clouds has been modelled, and appears possible (Tokano *et al.*, 2001b). Non-convective electrification is significant in Titan's chemistry and climate because most of the chemical reactions forming organic compounds are triggered by cosmic ray ionisation, and the stratospheric haze particles are probably electrically charged (Toon *et al.*, 1992; Fulchignoni *et al.*, 2002; Bakes *et al.,* 2002; Barth and Toon, 2003). The famous experiment illustrating the formation of amino acids by electrical discharge in a synthetic "primordial atmosphere" (Miller, 1953), similar to Titan's, suggests that atmospheric electrical studies of Titan may help to understand the evolution of life on Earth.

This provided a compelling motivation for the Huygens probe carried by the Cassini spacecraft, which entered Titan's atmosphere on 14 January 2005. One of the stated objectives of the Huygens Atmospheric Structure Instrument (HASI), carrying substantial electrical and meteorological instrumentation, is to measure atmospheric electrification (Fulchignoni *et al.*, 2002) The Surface Science Package (SSP) (Zarnecki *et al.*, 2002) will also contribute by making some meteorological measurements on its descent, and by determining the surface state and composition. This Section summarises the pre-Huygens understanding of Titan's atmospheric electrification, and gives an overview of the measurements to be carried out by Huygens.

### *6.1. Non-convective electrification*

Cosmic ray ionisation is important in the atmospheres of the outer planets and their satellites, because the intensity of solar UV radiation decreases with the square of the distance from the Sun (Bernard *et al.*, 2003). Titan's upper atmosphere becomes ionised by magnetospheric electrons from Saturn (when its orbit is within Saturn's magnetosphere) and solar UV radiation, but the electrons only contribute to ionisation at $z \geq 600$km, and atmospheric density prohibits penetration of the solar UV radiation below ~40km (Molina-Cuberos *et al.*, 1999a; 2002). Cosmic rays are therefore the dominant source of ionisation in Titan's middle and lower atmosphere. Titan has no geomagnetic field (see Table III) so there is no latitudinal variation of cosmic ray ionisation. The modulation of cosmic rays by the solar wind is reduced compared to planets closer to the Sun, with little ionisation variation over the solar cycle. According to Molina-Cuberos *et al.* (1999b) the ionisation rate is insensitive to solar cycle variations below 60km. Modelled ionisation rates for comparable regions on Earth and Titan are compared in Figure 7. A similar plot, showing ionisation rate variation with atmospheric density, is presented in Borucki *et al.* (1987). As the surface characteristics are unknown, but could be water ice or hydrocarbons (Lorenz, 2003), ion production from natural radioactivity at the surface is likely to be negligible. As on Earth, most of the cosmic ray ionisation near the surface is from muons produced by a cascade of sub-atomic particles from the decay of a high-energy primary cosmic ray in the upper atmosphere (O'Brien, 1971; Molina-Cuberos *et al.*, 1999b). Borucki *et al.* (1987) explain that most of the differences in ionisation rate between Earth and Titan result from the different density-altitude profile on Titan.

Although Titan's atmospheric ionisation rates are lower than on Earth, the atmospheric conductivity is probably ~300 times greater because of its different chemical composition (Borucki *et al.*, 1987). Titan lacks electrophilic species for electrons to attach to after ionisation, and mobile free electrons are expected to be abundant and dominate the air conductivity. The concentration of electrophilic species is not well-known, particularly for the lower atmosphere, but the negative ion concentration and conductivity were calculated by Borucki *et al.* (1987) and Molina-Cuberos *et al.* (2001b) for different mixing ratios of electrophilic species. The maximum expected mixing ratio $10^{-11}$ gives a negative ion concentration which is similar to the positive ion concentration, a few hundred $cm^{-3}$ up to 20km. A lower mixing ratio of $10^{-15}$ leads to <10 negative ions $cm^{-3}$. Much of Titan's chemistry is driven by cosmic ray ionisation, which has motivated modelling studies, first by Capone *et al.* (1980), followed by Borucki *et al.* (1987), and Molina-Cuberos *et al.* (1999a,b). Capone *et al.* (1980), and Borucki *et al.* (1987) predicted that nitrogenated cation clusters such as $NH_4^+(NH_3)_n$ and $HCNH^+(HCN)_n$ would be common, whereas Molina-Cuberos *et al.* (1999a) calculated that hydrocarbons were formed more efficiently, and ions like $CH_5^+CH_4$ would be abundant. Molina-Cuberos *et al.* (1999a) also excluded ammonia and related species from their calculations based on the premise that ammonia has not yet been detected in Titan's atmosphere, whereas the earlier models included ammonia as a possible atmospheric species. Since these modelling studies, Bernard *et al.* (2003) have detected ammonia in laboratory simulations of Titan's atmosphere; Huygens data, discussed in Section 6.3, should provide definitive information on the chemical composition of the Titan atmosphere.

The different ion species predicted in Titan's atmosphere all have similar masses (~100 amu) and total number concentrations, and therefore, from Equation (2), all predict a positive air conductivity of $10^{-15}$ $Sm^{-1}$ at the surface, rising to $10^{-11}$ $Sm^{-1}$ at 70km. The positive surface conductivity is similar to that on Earth, but the air conductivity increases more slowly with height than on Titan because of the shallower decrease in atmosphere density with altitude. Borucki *et al.* (1987) included aerosol charging in their model based upon the aerosol data from Voyager. The free electrons in the atmosphere readily attach to aerosols, causing a negatively charged aerosol population. Borucki *et al.* (1987) do not state the method used to compute the mean aerosol charge, but using equation 19 in Harrison and Carslaw (2003) gives similar results to those of Borucki *et al.* (1987), summarised in Table IV. The total conductivity was relatively insensitive to the aerosol concentration and size distribution below 100km altitude.

Calculations suggest that it is too cold for methane or ethane to nucleate onto charged particles by the Wilson mechanism (Section 3.3). Whilst there is some uncertainty over which parts of Titan's atmosphere are supersaturated, the supersaturations required for ion-induced nucleation appear unfeasibly high in the cold cloud-forming regions. Navarro-Gonzalez and Ramirez (1997) suggest that methane clouds have charged nuclei, but it is unclear if this is simply based upon the assumption that, as on Earth, an aerosol particle on Titan is statistically likely to be charged (Borucki *et al.*, 1987; Navarro-Gonzalez and Ramirez, 1997). Electrical charge controls the particle size and optical depth of the thick stratospheric haze layer, and therefore has to be included in Titan aerosol and cloud microphysics models (Toon *et al.,* 1992; Barth and Toon, 2003). Toon *et al.* (1992) compared the charging required to produce the observed particle characteristics with the aerosol charging predicted by Borucki *et al.* (1987), and found some inconsistencies. The particles were required to have three times more charge than electrical models alone predicted to explain the polarisation observations. This could be caused by the assumption that the particles were spherical, or inaccuracies in the Borucki *et al.* (1987) model. The electrical, chemical composition and aerosol observations to be supplied by Huygens should permit calculation of the charged particle profile in Titan's atmosphere.

*6.2. Convective electrification*

Voyager 1 did not detect lightning during its 1980 fly-by of Titan, implying that lightning could be rare, not detectable by Voyager 1 (Desch and Kaiser, 1990), or that it simply does not exist on Titan. Desch and Kaiser (1990) used their results to determine a minimum energy threshold of ≤1 MJ/flash, but recent electrostatic calculations (Fischer *et al.*, 2004) indicate that average Titan lightning strokes could dissipate 0.1-10 GJ/flash. Conversely, as the Cassini RPWS instrument is much more sensitive than the Voyager radio detectors, its failure to detect lightning on Titan indicates that it could be weaker than 1MJ/flash. As Fischer *et al.* (2004) emphasise, a lightning energy distribution, as for terrestrial thunderstorms, is quite likely, and the observations remain inconclusive. If Titan lightning does exist, the characteristic ELF signals described in Section 2.1 are predicted between 11 and 15Hz, which may be detectable by Huygens but probably not by Cassini (Morente *et al.*, 2003). Lightning is often suggested as a possible trigger for some of the complicated organic chemistry in Titan's atmosphere. A detailed discussion of Titan's meteorology is beyond the scope of this paper, but its lower atmosphere contains transient methane clouds, first detected from Earth (Griffith *et al.*, 1998), and recently by Cassini (Porco *et al.*, 2005b). Methane is sometimes suggested to be an unlikely candidate for lightning because of its low dc permittivity, 1.7 compared to 80 for water. However, in a detailed model of convection and charging mechanisms, Tokano *et al.* (2001b) found that electric fields of up to 2 $MVm^{-1}$ could be developed in methane clouds, which could trigger cloud-to-ground lightning. This model, and Titan lightning in general, are reviewed in Desch *et al.* (2002). If Titan's methane rain evaporates before it reaches the surface, this could change the currents carried to the ground by precipitation. Until more data on Titan's weather are available, it is difficult to estimate any more atmospheric electrical parameters.

*6.3. Huygens instrumentation*

The Huygens instrument contains six instrumentation packages, which will give a detailed picture of Titan's atmosphere and surface conditions at the landing site. Atmospheric electrical and meteorological parameters, aerosol properties, atmospheric composition, wind conditions, and surface physical properties will all be measured. The HASI will detect acceleration, atmospheric pressure, temperature, electrical conductivity, dc atmospheric electric field, lightning, acoustic noise and radar echoes (Fulchignoni *et al.*, 2002). The PWA subsystem contains the electrical instrumentation (Molina-Cuberos *et al.*, 2001b), including :

- two relaxation probes measuring conductivity in the range $10^{-14}$ Sm$^{-1}$-$10^{-12}$ Sm$^{-1}$ and DC electric field,
- two AC impedance sensors measuring conductivity in the range $10^{-14}$ Sm$^{-1}$ - $10^{-7}$ Sm$^{-1}$, and
- an acoustic sensor to detect noise from turbulence, thunderstorms or even methane raindrops (Desch *et al.*, 2002).

Models of HASI have been tested in the terrestrial atmosphere on several occasions and performed well each time (López-Marino *et al.*, 2001; Fulchignoni *et al.*, 2004). The HASI is the first space probe to carry instrumentation to detect non-convective atmospheric electrification, and will soon give the first detailed picture of an extraterrestrial atmospheric electrical environment[2]. Data from other Huygens instruments could be used to deduce more subtle electrical properties. For example, molecular mass profiles could be used for estimating mean ion mobilities. Parameters such as the ion-aerosol attachment coefficient $\beta$, which is significant for investigation of ion-aerosol interactions (Equation (3)), could be determined based on aerosol size distribution measurements. It would be tantalising to consider the possibility of particle formation by indirect ion-mediated nucleation in Titan's atmosphere, as may happen in the terrestrial atmosphere, given the similarities between Titan's and Earth's electrical environments. Supersaturations appropriate for ion-induced nucleation are unlikely on both Titan and Earth, but some of the organics in Titan's atmosphere could have a low enough vapour pressure to assist nucleation, as is thought to occur on Earth. Huygens data may provide enough information about the atmospheric structure and composition for complex ion-aerosol models like the terrestrial one developed by Yu and Turco (2001) to be applied to Titan. Ultimately, this information could be used to help terrestrial atmospheric electrical researchers. In particular the relative stability of the Titan atmospheric electrical system could be useful when compared to the extreme temporal and spatial variability facing terrestrial researchers in "fair-weather" atmosphere electricity. Huygens should also yield important information on the role of atmospheric electricity in the prebiotic terrestrial environment.

[2] This footnote will be used to briefly mention any Huygens results that are published at the time the paper goes to press.

## 7. Uranus and Neptune

Uranus and Neptune are broadly similar in composition and structure to Jupiter and Saturn (Section 5). Their characteristic marine colours are a consequence of methane (the third most abundant species, after hydrogen and helium) absorbing the red and yellow part of the spectrum which dominates the colours of Jupiter and Saturn. Since the Voyager 2 flybys in 1986 and 1989 respectively, new data have come from adaptive optics observations from ground-based telescopes, and from the Hubble Space Telescope. Relatively little is known about the atmospheres of Uranus and Neptune compared to the closer gas giant planets, but both have active zonal winds and changing cloud systems.

### *7.1. Uranus*

Uranus is unique amongst the gas giant planets for two reasons. First, it lies almost in the plane of the ecliptic, nearly perpendicular to the other planets, so its seasons are very severe with the summer hemisphere almost completely vertically illuminated by sunlight. Secondly, it has no internal heat source, so its convective activity is much lower than that of the other gas giants (Miner, 1998). This accounts for the similar temperatures of Uranus and Neptune, even though Uranus is much closer to the Sun, and the relative calmness of the Uranian atmosphere compared to Jupiter, Saturn and Neptune. Although lightning might seem unlikely in this environment, suggestive radio emissions (Uranian electrostatic discharges, UEDs) were observed by Voyager 2 (Zarka and Pedersen, 1986). CO has recently been detected in the Uranian atmosphere which could have been produced by lightning (Encrenaz *et al.*, 2004). The radio observations were similar to SEDs detected by Voyager 2 and Cassini. Comparison with the new Cassini observations, discussed in Section 5, could provide more information on the causes of the UEDs.

Other than hydrogen, helium and methane, the Uranian troposphere contains carbon dioxide, phosphine and stratospheric unsaturated triple bonded hydrocarbons such as acetylene ($C_2H_2$) and diacetylene ($C_4H_2$). Ethane ($C_2H_6$) ice clouds exist at ~1000hPa and there are probably hydrogen sulphide ($H_2S$) clouds at 3100hPa, possibly with water ice clouds below (Miner, 1998; Encrenaz *et al.*, 2004) The hydrocarbon chemistry is triggered by solar UV radiation resulting in haze layers, probably of solid ethane, acetylene and diacetylene at ~0.05-0.13 bar. Capone *et al.* (1979) predicted that the dominant ion species produced by cosmic rays in the Uranian troposphere and stratosphere were $C_3H_{11}^+$ and $C_2H_9^+$. Cosmic rays are likely to be the only ionisation at $p$>100hPa where solar UV radiation does not penetrate. Moses *et al.* (1992) stated that the assumptions used in the Capone *et al.* (1979) model were out of date, but there still appear to have been no more studies of cosmic ray ionisation in Uranus' lower atmosphere, despite new theoretical work on upper atmosphere ionisation in the gas giants (Velinov *et al.*, 2004).

### *7.2. Neptune*

As indicated in Section 7.1, Neptune's atmospheric convection makes it seem more likely to have lightning than Uranus. Observations show that Neptune is a more convective planet than Uranus (but less so than Jupiter and Saturn) with several cloud layers, of methane, $H_2S$-ammonia ($NH_3$), water and $NH_4SH$ (ammonium hydrosulphide), and a stratospheric haze layer (Gibbard *et al.*, 1999). Recent ground-based observations suggest that convection transports methane ice clouds upwards to the tropopause, and there is subsidence of the stratospheric haze in other locations causing a global haze cycle (Gibbard *et al.*, 2003). As on Uranus, the haze is composed of photochemically formed complex hydrocarbons, with higher order alkanes like propane ($C_3H_8$) and butane ($C_4H_{10}$) (Moses *et al.*, 1992). A search for lightning by Voyager 2 revealed four possible radio emissions, and sixteen whistler events detected by two different instruments (Gibbard *et al.*, 1999). This is more evidence for Neptunian lightning than was detected for Uranus; however, lightning could not be detected optically. It was speculated that the discharges were too deep in the atmosphere to be detected at optical wavelengths (Borucki and Pham, 1992). Microphysical modelling found that lightning was most likely to occur in the ammonium hydrosulphide clouds, but the study was limited by a lack of experimental data on the electrical properties of ammonium hydrosulphide, particularly the efficacy of charge transfer (Gibbard *et al.*, 1999). More ground-based laboratory investigations could improve our understanding of the charging mechanisms involved.

Capone *et al.* (1977) calculated that the ion species formed by cosmic rays were very similar to those on Uranus, dominated by $C_3H_{11}^+$ and $C_2H_9^+$. Inspired by reports of ion-induced nucleation in the terrestrial atmosphere, Moses *et al.* (1989) proposed a similar mechanism to explain observations of Neptune's albedo varying in antiphase with the 11-year solar cycle (Lockwood and Thompson, 1986). Ions formed by cosmic rays, based on the Capone *et al.* (1977) model, were hypothesized to produce a methane ice haze varying with the ion production rate, but an alternative mechanism was that the colour of Neptune's aerosols could be affected by the solar UV flux (Baines and Smith, 1990).

After the Voyager flyby an updated and more detailed model could be produced (Moses *et al.*, 1992) which compared the relative efficiencies of homogeneous nucleation (condensation of gas onto a nucleus of the same substance), heterogeneous nucleation and ion-induced nucleation. The low temperatures substantially limited kinetic nucleation mechanisms and increased the relative importance of electrical processes. Ion-induced nucleation was only expected to be more efficient than heterogeneous nucleation for "heavy and sluggish" molecules with low vapour pressures, such as diacetylene.

Lockwood and Thompson (2002) continued to monitor Neptune's albedo and found that a steady rising trend dominated the 11-year variation since about 1990. They correlated the Neptune albedo with the 121.6nm Lyman-$\alpha$ flux, and the terrestrial neutron flux as proxies for the Neptune UV and cosmic ray fluxes, respectively. A statistically significant correlation was only observed between the albedo and Lyman-$\alpha$ flux. Lockwood and Thompson (2002) acknowledged the incompleteness of a correlation analysis in determining mechanisms, but implied that the UV mechanism is more probable. Even this is not straightforward, since it is very hard to explain *any* solar effects on the Neptune atmosphere given its great distance from the Sun. Further analysis would require improved cloud microphysics, cosmic ray ionisation and photochemical models. It would also be interesting to apply the Moses *et al.* (1992) ion-induced nucleation model to Uranus, which has similar thermal and chemical properties, to compare ion-induced nucleation with other cloud formation mechanisms.

## 8. Triton and Pluto

Triton, Neptune's largest moon and Pluto, the ninth planet, have similarly tenuous and cold atmospheres with surface pressures of <1 Pa. Both these worlds initially seem cold and barren, but the lack of solar energy input may lead to a greater role for electrical forces. The atmospheres of these small objects (Pluto is smaller than the Earth's Moon, and Triton smaller still) are fundamentally different from any other planets. For most of their orbital periods, their atmospheres are frozen, but solar heating warms their solid nitrogen surfaces (Hubbard, 2003). A similar physical process forms the coma of a comet but, unlike comets, Pluto and Triton are just massive enough to retain the emitted gas as an atmosphere.

Triton is the better understood of the two bodies, as it was visited by Voyager 2 in 1989, when its atmosphere was discovered. It also yielded unexpected geological and meteorological activity with observations of geysers and clouds (Delitsky *et al.* 1990; Soderblom *et al.*, 1990). Ion-induced nucleation is the only existing explanation for Triton's thin haze, predicted to form by condensation of nitrogen onto ions produced either by cosmic rays or magnetospheric particles (Delitsky *et al.,* 1990). In Triton's cold atmosphere, larger nitrogen clusters are more stable, which encourages ion-induced nucleation once supersaturation has been reached. Delitsky *et al.* (1990) showed that high supersaturations of ~10, and associated nucleation rate increases, could be achieved by a relatively small drop in temperature because of the sensitivity of nitrogen's vapour pressure to temperature in the low-temperature regime. The coldest region, and therefore the most favourable for particle formation onto nitrogen ion clusters, is the tropopause at 9 km, where the haze is observed (Delitsky *et al.,* 1990; Soderblom *et al.*, 1990).

The "cometary" model of Pluto and Triton's atmosphere generation is supported by observations of rapid seasonal change. Triton's atmospheric pressure doubled from 1990-1998, and Pluto's atmospheric pressure has also doubled over the last 14 years (Elliot *et al.*, 1998; Sicardy *et al.*, 2003). The warming observed was consistent with nitrogen cycle models, although Pluto was moving away from the Sun during this time. Occultation observations implied dynamical activity in Pluto's atmosphere, boundary layer effects and the presence of a morning haze layer. Elliot *et al.* (2003) also suggested that light extinction observed on Pluto could be related to the presence of photochemically produced aerosol particles. If this haze layer does exist, it is reasonable to expect comparable mechanisms to act on Pluto and Triton. A mission to Pluto is needed to obtain further information; Hubbard (2003) describes a NASA New Horizons mission planned for launch in 2006 with a Pluto flyby about 10 years later. It is not known exactly when Pluto's atmosphere will next freeze, but this constrains the timing of any space mission. It is uncertain whether the timescales and funding required for the New Horizons mission are compatible with this scientific need.

## 9. Conclusions

The planetary atmospheric electrical systems discussed in this paper fall naturally into three groups, similar but not identical to the traditional classifications of terrestrial, gas giant and outer planets. The "terrestrial" planetary atmospheric electric systems are, like the terrestrial planets, those which are most similar to Earth. These planets have well-defined surfaces and charged particle populations, with a high probability of electrostatic discharge. They are consequently the most likely group to fulfill the conditions for a global electric circuit like Earth's. The criteria for a global circuit defined in Section 1, and the evidence for each of them in the planetary atmospheres discussed are summarised in Table V. The classical taxonomy includes Earth, Venus and Mars but the atmospheric electrification classification scheme must also include the satellite, Titan, which has similar atmospheric pressure and composition to Earth. Active non-convective electrification has been predicted for Venus, but the existence of lightning is uncertain. A Martian global circuit has been proposed; it is comparable to the terrestrial model, but driven by dust storms with the opposite dipole structure to terrestrial thunderclouds. Experimental corroboration of these predictions has been limited by the complete absence, to date, of any *in situ* electrical measurements in extraterrestrial atmospheres. The Huygens probe carries substantial atmospheric electrical instrumentation and should produce enough data to validate the numerous predictions of the electrical properties of Titan's atmosphere and solve some of the outstanding issues, such as whether lightning, and electrophilic species, exist.

The gas giant planets (Jupiter, Saturn, Uranus and Neptune) all have lightning and active weather systems. It has been suggested that the condensation of gases onto ions, ion-induced nucleation, could be relevant, particularly in the atmosphere of Neptune, and possibly also of Uranus. Despite the existence of ions, aerosol, polar molecules and convection, it is difficult to apply the terrestrial model of a global circuit to these planets because of the probable absence of a conducting surface. A different electrical model for the gas giant planetary atmospheres may be appropriate, but further theoretical work is required first.

Both the atmospheres of the small outer solar system bodies, Pluto and Neptune's satellite Triton, develop seasonally. Triton benefited from a 1989 visit by Voyager 2, in which its atmosphere and thin tropospheric haze layer were detected, but studies of both bodies have been complicated by a lack of observational data. Pluto's atmosphere was only discovered in the 1980s from telescope occultation observations and, with great ingenuity, the existence of atmospheric aerosols has recently been deduced from ground-based telescope studies, possibly as a haze layer. Triton's haze layer is expected to have been formed by ion-induced nucleation, and similar mechanisms could be possible on Pluto.

The same atmospheric electrical processes act across the solar system. Lightning has been detected on many other planets, and Venus, Mars and Titan could have quasi-terrestrial global circuits. Jupiter and Saturn's electrical systems are poorly understood and may have to await further understanding of the nature of their high-pressure interiors. Ion-induced nucleation could occur on Venus, Neptune, Triton and possibly Uranus and Pluto. Ion-mediated nucleation has been observed on Earth, but there have not yet been any attempts to predict or identify it in other planetary atmospheres. In the cold, yet rapidly evolving, atmospheres of Triton and Pluto, there are so few forces acting that electric charge may be relatively important. Clear needs for further research can be identified based on a comparative approach to the themes discussed in this paper.

- *Direct measurements*: Electrostatic measurements on Mars are urgently required to characterise an environment that may soon be visited by humans. The role of charged particles in atmospheric chemistry and meteorology could be investigated simply with the minimum of instrumentation - one relaxation probe - to measure atmospheric conductivity and electric fields on future planetary atmosphere missions.
- *Application of terrestrial atmospheric physics*: Though there is still a paucity of terrestrial experimental atmospheric electrification data, model predictions are more easily tested on Earth. Models validated for Earth's atmosphere, like the ones used to predict ion-mediated nucleation, could be redeveloped to enhance our understanding of other electrically similar worlds like Titan and Venus. Ground-based laboratory experiments to investigate terrestrial thundercloud charging could be extended to study the electrical properties of charge-separating species in other atmospheres.
- *Development of planetary atmosphere theoretical work*: Excluding Titan and Mars, there has been little recent theoretical work on planetary atmospheric electrification. Detailed predictions of atmospheric electrification on Venus were last made over twenty years ago and would benefit from updating with modern observations and models. The theoretical basis for predicting ion-induced nucleation on Neptune and Triton could be usefully extended to investigate Uranus and Pluto, respectively.

The most exciting forthcoming development in the study of planetary atmospheric electrification is the Huygens probe. It should provide the most detailed observations ever made of the electrical environment on another planet. Using the Huygens data, the electrical contributions to the origins of life on Earth may become clearer.

**Table Captions**

Table I. Summary of electrification in planetary atmospheres. Atmospheric constituents, surface temperature and pressure are from Lewis (1997).

Table II. Sulphuric acid supersaturation in Venus' atmosphere, from Kolodner and Steffes (1998). Supersaturations required for condensation of sulphuric acid vapour onto charged particles are also shown. Temperatures are interpolated from Borucki *et al.* (1982).

Table III. Comparison of planetary magnetic fields, compiled from Stevenson (2003), Ip *et al.* (2000) and Smoluchowski (1979).

Table IV. Ion-aerosol properties at sea level for Earth, Titan and Venus.

Table V. Summary of global circuit criteria met by each planetary atmosphere.

**Figure Captions**

Figure 1. Conceptual diagram of the terrestrial global circuit, from Rycroft *et al.* (2000), reproduced with permission from Elsevier. Left hand side: disturbed weather, representing global thunderstorm activity. Right hand side: Fair weather. Arrows represent the direction of current flow.

Figure 2 Results from the Borucki *et al.* (1982) ion-aerosol model for Venus' atmosphere (reproduced with permission from Elsevier) a) ionisation rate profile; the solid curve shows predicted ionisation rate when the effects of muons are included, and the dashed line shows the result when the muon flux is ignored. b) ion and electron mobility profile; bipolar ion mobilities are almost the same and are plotted on the lower scale, whereas the electron mobility is plotted on the upper scale. c) ion and electron

concentration profile. Solid line: electrons, dashed line: positive ions, dot-dash line: negative ions, dotted line: results for an atmosphere where attachment to cloud particles is ignored. d) conductivity profile; solid line: total conductivity, dotted line: total conductivity assuming no aerosols, dashed line: positive conductivity, dot-dash line: negative conductivity, dot-dot-dash: electron conductivity

Figure 3. Abundance profile of gaseous sulphuric acid, showing supersaturation below the lower cloud in the Venusian atmosphere at 88°S. Retrieved from the S-band absorptivity profile of the Magellan orbit 6370 radio occultation experiment. Reproduced with permission from Elsevier from Kolodner and Steffes (1998).

Figure 4. (a) Supersaturations required for Wilson nucleation onto ions with 1, 2 and 5 electronic charges for the region at 43km shown in Figure 3 (b) Steady state charge distribution on Venusian aerosols at 40km, assuming a temperature of 395K and a mean aerosol radius of 0.25μm.

Figure 5. (a) Diurnal statistics of convective vortices seen by the Viking 2 lander, after Ringrose *et al.* (2003). b) Onset locations of large Martian dust storms, binned in 30º bands of longitude, after Newman *et al.* (2002). The time difference from the time at 0º is also indicated.

Figure 6. (a) Schematic of a possible Martian global electric circuit, based on Farrell and Desch (2001). (b) "Carnegie" diurnal electric field variation for Martian "fair weather", compared to the terrestrial Carnegie curve, plotted from the data in Harrison (2005).

Figure 7 Ionisation rates calculated to 45km for a nominal Titan atmosphere (100% $N_2$) and a modelled terrestrial atmosphere. The values for Titan are from Molina-Cuberos *et*

*al.* (1999b) and the values for Earth are calculated for mid-latitudes at the middle of the solar cycle, after Aplin and McPheat (2005).

| **Planet/ *Moon*** | **Principal atmospheric constituents (% by vol)** | **Surface p (ratio to Earth), T(K)** | **Atmospheric electrical processes** | | | **Main Refs** |
|---|---|---|---|---|---|---|
| | | | **Lightning** | **Charged aerosol** | **Ion nucleation?** | |
| **Venus** | $CO_2$ 96.5<br>$N_2$ 3.5 | 92, 700 | Probably not, but could be sprite-like | Almost certain, not yet observed | Possibility of direct "Wilson" nucleation below clouds | Borucki *et al.* (1982) |
| **Earth** | $N_2$ 78.0<br>$O_2$ 20.9 | 1,288 | Cloud-to-ground, intracloud, intercloud, sprites | Yes | No "Wilson" nucleation but probably ion mediated nucleation | Harrison and Carslaw (2003), Rycroft *et al.* (2000) |
| **Mars** | $CO_2$ 95.3<br>$N_2$ 2.7<br>Ar 1.6<br>$O_2$ 0.13<br>CO 0.08 | $5 \cdot 10^{-3}$, 210 | Never detected, but electrostatic dust discharges are likely | Triboelectrically charged dust | Unlikely: no supersaturation or condensable species | Farrell and Desch (2001), Fillingim (1998) |
| **Jupiter** | $H_2$ 89.9<br>He 10.2 | Gas giant | Detected optically | Likely | Possible, but unlikely to be significant | Capone *et al.* (1979), Lewis (1997) |
| **Saturn** | $H_2$ 96.3<br>He 3.3 | Gas giant | Radio emissions detected by Voyager and Cassini | Likely | Similar to Jupiter? | Kaiser *et al.*, (1983), Gurnett *et al.*, (2005) |
| ***Titan*** | $N_2$ 94<br>$CH_4$ 6<br>$H_2$ 0.2<br>CO 0.01 | 1.5, 94 | Predicted by models; awaiting Huygens results | Almost certain | Probably too cold for "Wilson" nucleation; awaiting Huygens observations | Molina-Cuberos *et al.* (1999a,b), Tokano *et al.* (2001a) |
| **Uranus** | $H_2$ 82.5<br>He 15.2<br>$CH_4$ 2.3 | Gas giant | Possible; suggestive radio signals from Voyager | Likely: limited by abundance of neutral aerosols | Similar to Neptune? | Miner (1998) |
| **Neptune** | $H_2$ 80.0<br>He 19.0<br>$CH_4$ 1.5 | Gas giant | Possible; whistlers detected by Voyager | Likely: limited by abundance of neutral aerosols | Wilson nucleation of diacetylene possible | Moses *et al.* (1992) |
| ***Triton*** | $N_2$ 99.99<br>$CH_4$ 0.01 | $10^{-5}$, 37 | Highly unlikely, too cold | Possible; haze could be charged | Wilson nucleation predicted | Delitsky *et al.* (1990) |
| **Pluto** | $CH_4$<br>$N_2$ | $3 \times 10^{-6}$, 50 | Highly unlikely, too cold | Similar to Triton? | Similar to Triton? | Hubbard (2003) |

**Table I**

| **Figure in Kolodner and Steffes (1998)** | **Latitude** | **Minimum height for supersaturation (km)** | **Estimated temperature (K)** | **Approximate supersaturation required for condensation onto particles with *qe* electronic charges (%)** | | |
|---|---|---|---|---|---|---|
| | | | | ***qe*=1** | ***qe*=2** | ***qe*=5** |
| Fig 7 | 67ºN | 47 | 372 | >100 | >10 | 3 |
| Fig 8 | 67ºN | 46.5 | 394 | >10 | 7 | 0.2 |
| Fig 9 | 67ºN | 45.5 | 381 | 100 | 10 | 2 |
| Fig 10 | 88ºS | 42.5 | 401 | >10 | 5.5 | 1 |
| Fig 11 | 88ºS | 41.7 | 406 | >10 | 5 | <1 |
| Fig 12 | 0º | 44.5 | 388 | >10 | 10 | 1.5 |

**Table II**

| **Planet/*Moon*** | **Approximate surface magnetic field, Tesla (T)** |
|---|---|
| Venus | $< 10^{-8}$ |
| Earth | $5 \times 10^{-5}$ |
| Mars | $10^{-9} - 10^{-4}$ |
| Jupiter | $4.2 \times 10^{-4}$ |
| Saturn | $2 \times 10^{-5}$ |
| *Titan* | $< 10^{-7}$ |
| Uranus | $2 \times 10^{-5}$ |
| Neptune | $2 \times 10^{-5}$ |
| *Triton* | Probably negligible |
| Pluto | Not known |

**Table III**

| **Typical ion and aerosol properties at sea level** | | **Earth** | **Titan** | **Venus** |
|---|---|---|---|---|
| Small ion concentration ($cm^{-3}$) | $n_+$<br>$n_-$ | 200-2000, depending on natural radioactivity and aerosol | 500<br>1-100, depending on number of electrophilic species | ~500 |
| Small ion/electron mobility ($Vm^{-1}s^{-2}$) | $\mu_+$<br>$\mu_-$ | 1.2 x$10^{-4}$<br>1.9 x$10^{-4}$ | 1.2 x$10^{-4}$<br>depends on existence of electrophilic species | ~5 x $10^{-6}$, positive ions slightly more mobile |
| Bipolar atmospheric conductivity ($Sm^{-1}$) | $\sigma_+$<br>$\sigma_-$ | ~$10^{-14}$ | $10^{-15}$<br>$10^{-10}$ | ~5 x $10^{-16}$ |
| Space charge ($Cm^{-3}$) | $\rho$ | 4x$10^{-12}$ | 2x$10^{-12}$? | Surface properties not known well enough |
| Typical mean aerosol charge (number of electronic charges) | $j$ | -1, assuming mean aerosol radius = 0.2µm | - 5 to 50, depending on mean aerosol size | +0.2, assuming mean aerosol radius = 0.125µm |

**Table IV**

| **Planet/ *Moon*** | **Polar molecules** | **Charge separation** | **Conducting upper and lower layer?** | **Mobile charged particles in lower atmosphere** |
|---|---|---|---|---|
| **Venus** | Yes | Possible | Yes | Yes |
| **Earth** | Yes | Yes | Yes | Yes |
| **Mars** | Yes | Probable, but not observed | Upper: yes Lower: doubtful | Yes |
| **Jupiter** | Yes | Yes | Upper: yes Lower: doubtful | Probably not in deep atmosphere |
| **Saturn** | Yes | Yes | Upper: yes Lower: doubtful | Probably not in deep atmosphere |
| ***Titan*** | Yes | Possible, but not observed | Upper: yes Lower: not known | Probably |
| **Uranus** | Yes | Yes | Upper: yes Lower: doubtful | Probably not in deep atmosphere |
| **Neptune** | Yes | Yes | Upper: yes Lower: doubtful | Probably not in deep atmosphere |
| ***Triton*** | Not detected | Probably not | Not known | Probably |
| **Pluto** | Not detected | Probably not | Not known | Probably |

**Table V**

**Figures**

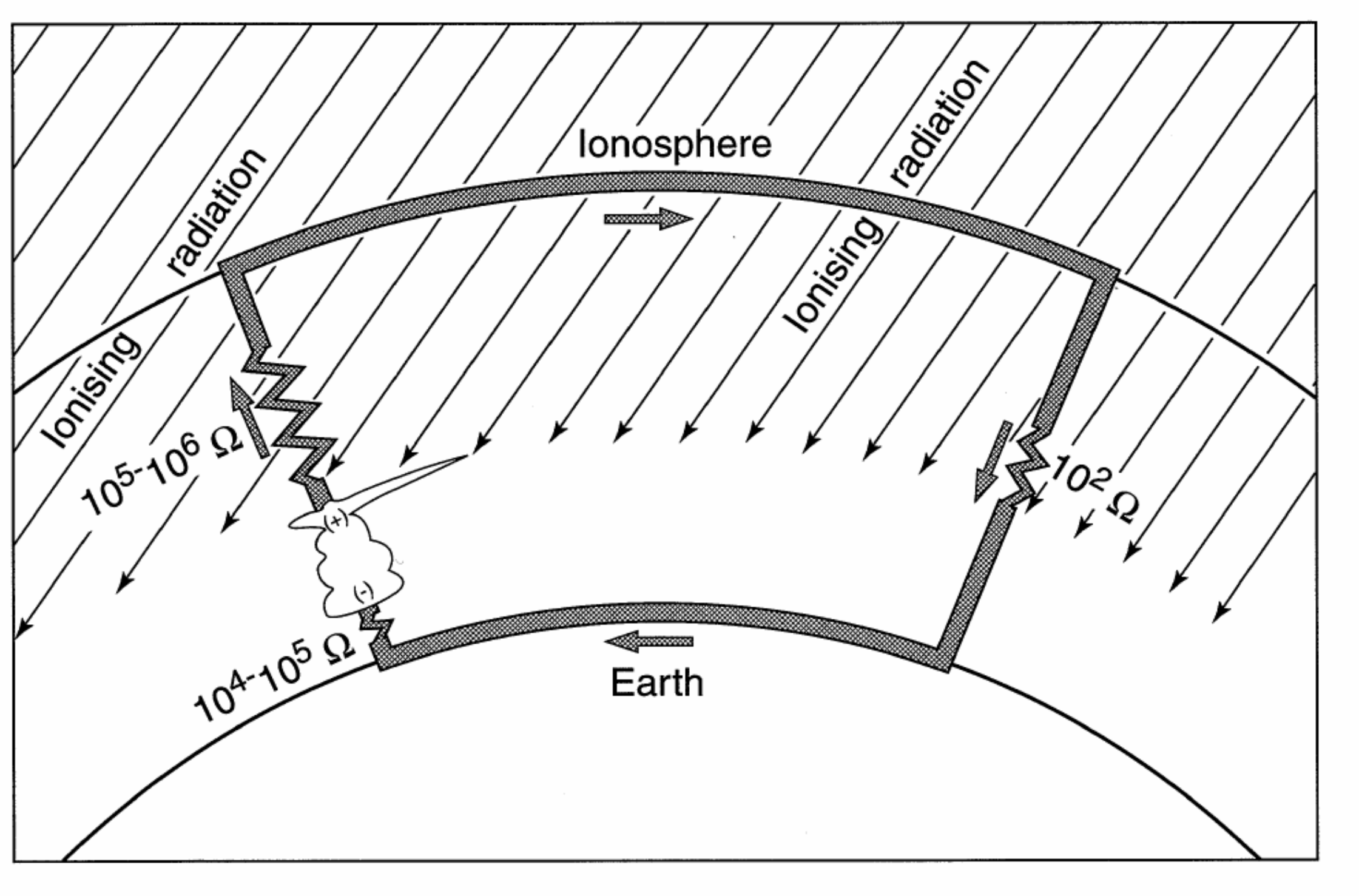

**Figure 1**

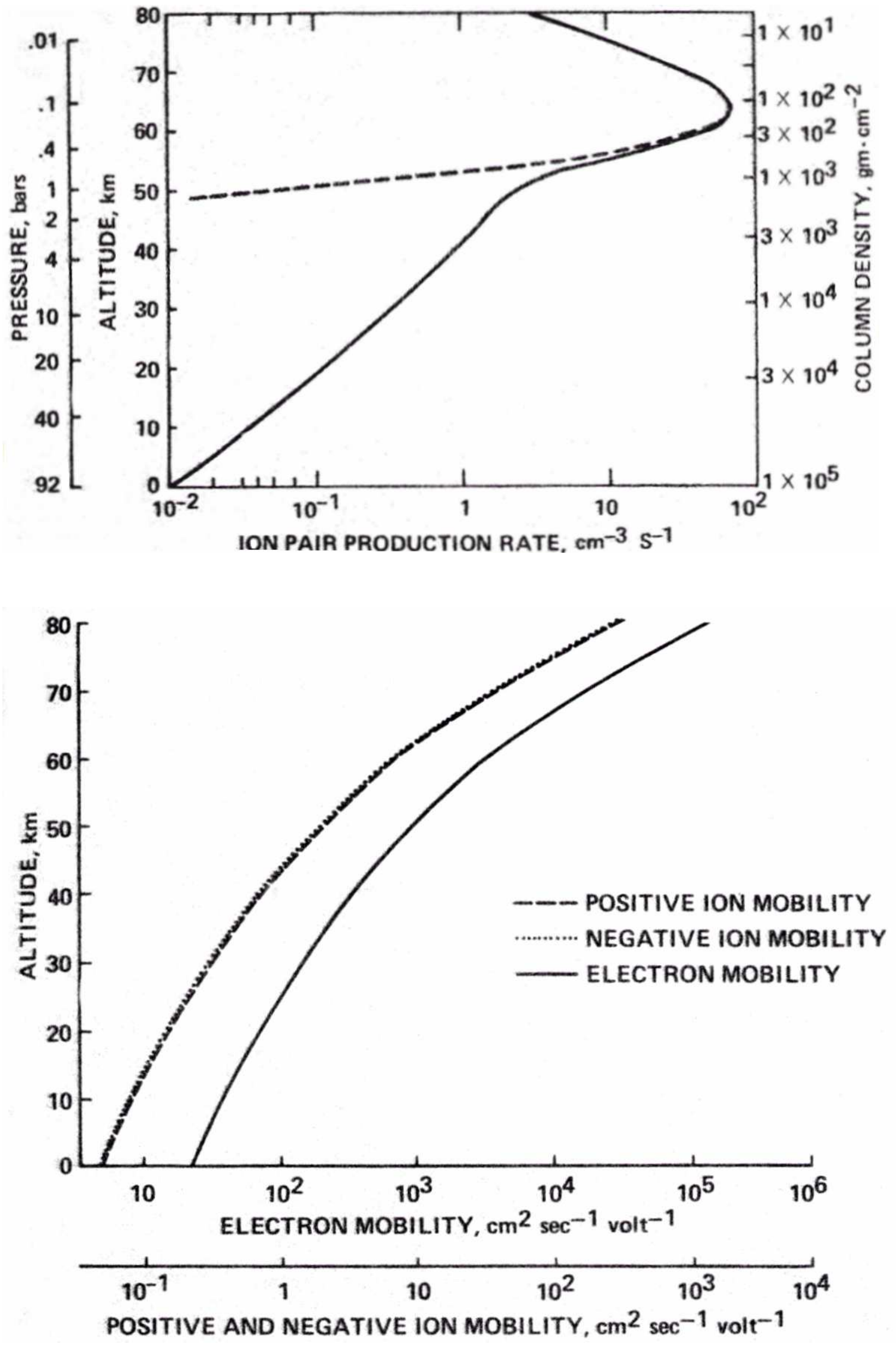

PRESSURE, bars
.01
.1
.4
1
2
4
10
20
40
92
ALTITUDE, km
80
70
60
50
40
30
20
10
0
$10^{-2}$
$10^{-1}$
1
10
$10^2$
ION PAIR PRODUCTION RATE, $cm^{-3}$ $S^{-1}$
COLUMN DENSITY, $gm \cdot cm^{-2}$
$1 \times 10^1$
$1 \times 10^2$
$3 \times 10^2$
$1 \times 10^3$
$3 \times 10^3$
$1 \times 10^4$
$3 \times 10^4$
$1 \times 10^5$
ALTITUDE, km
POSITIVE ION MOBILITY
NEGATIVE ION MOBILITY
ELECTRON MOBILITY
10
$10^2$
$10^3$
$10^4$
$10^5$
$10^6$
ELECTRON MOBILITY, $cm^2$ $sec^{-1}$ $volt^{-1}$
$10^{-1}$
1
10
$10^2$
$10^3$
$10^4$
POSITIVE AND NEGATIVE ION MOBILITY, $cm^2$ $sec^{-1}$ $volt^{-1}$

(a)

(b)

**Figure 2**

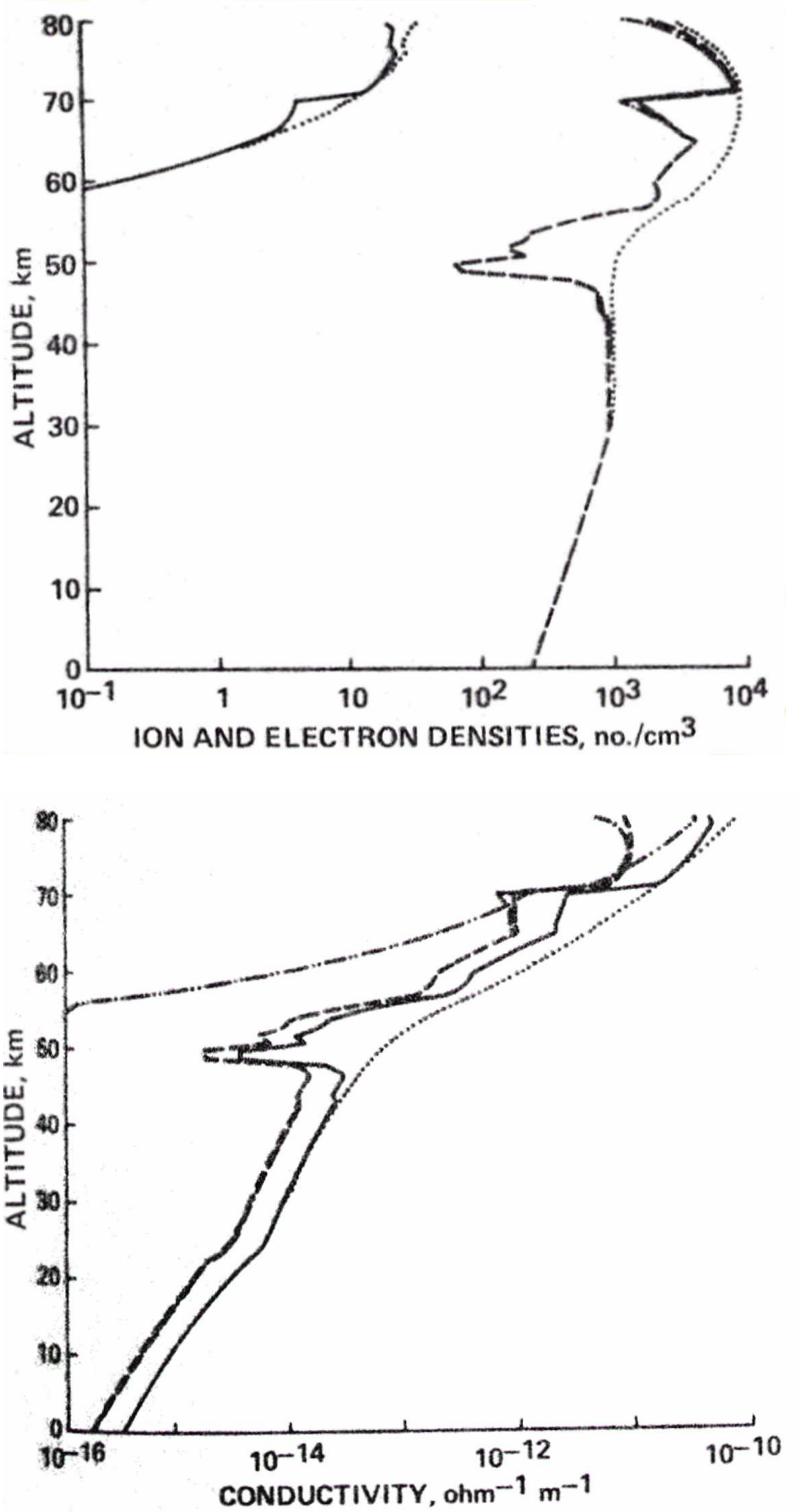

(c)

(d)

**Figure 2**

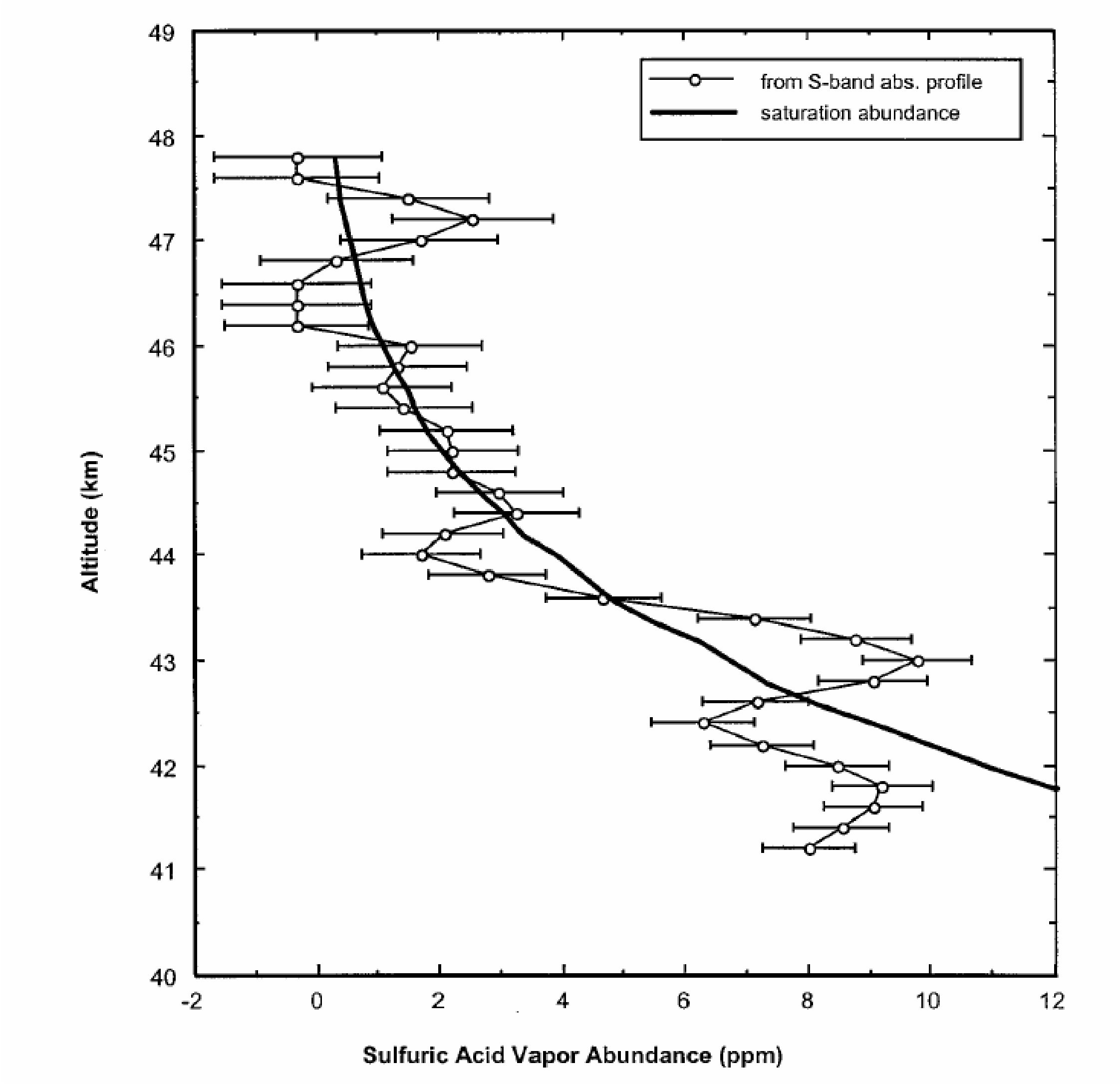

**Figure 3**

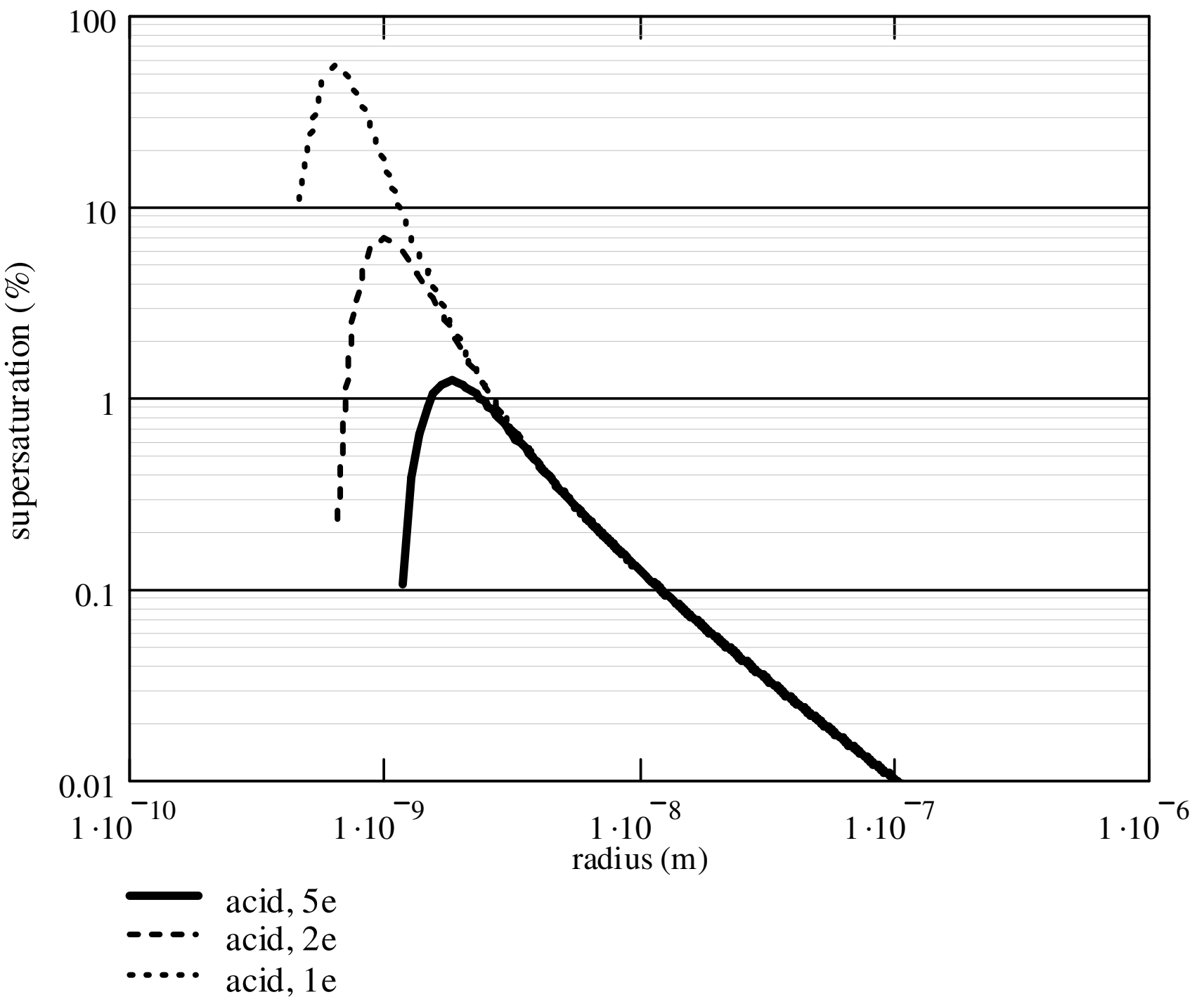

(a)

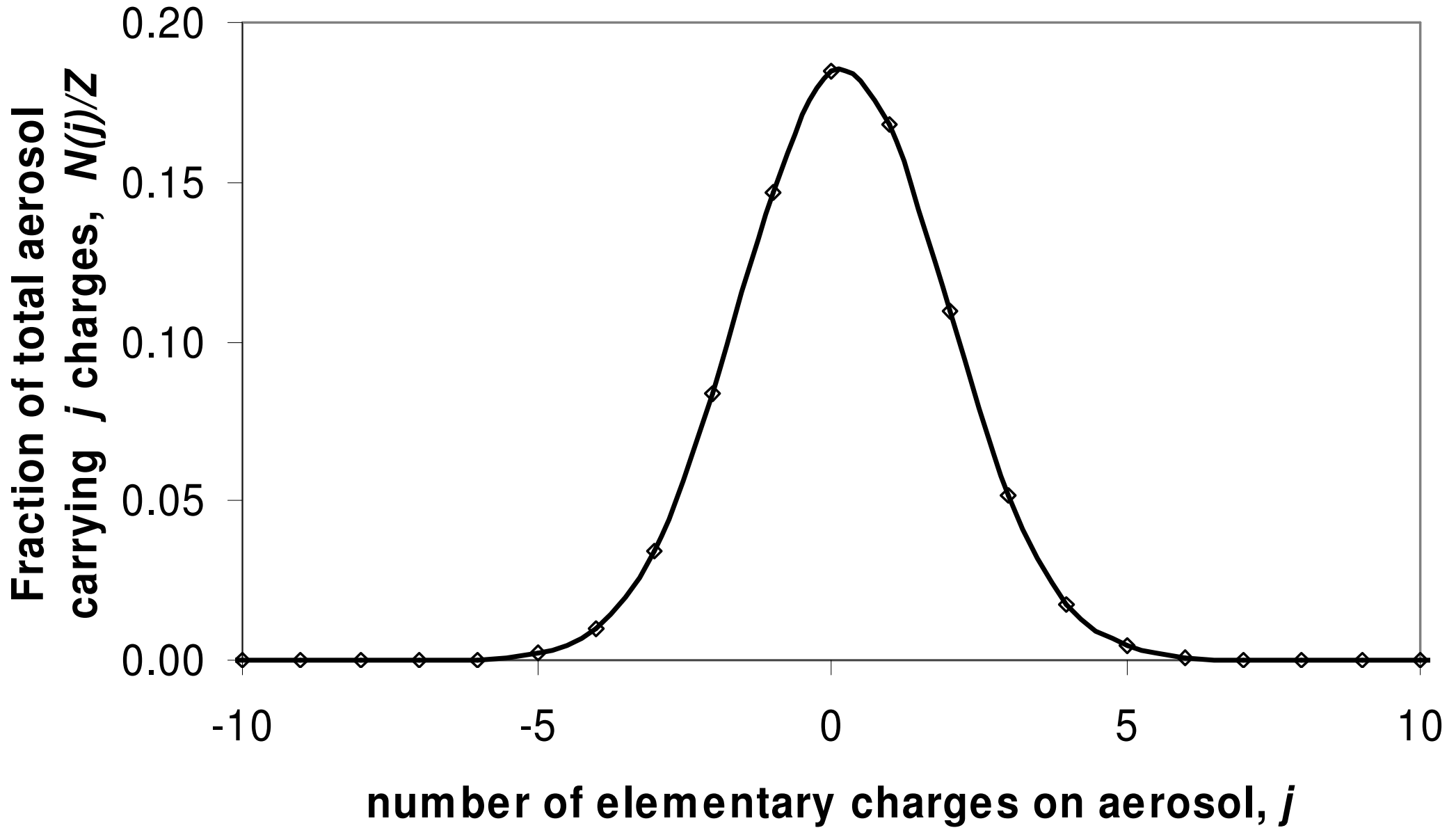
0.20
0.15
0.10
0.05
0.00
Fraction of total aerosol carrying j charges, N(j)/Z
-10
-5
0
5
10
number of elementary charges on aerosol, j

(b)

**Figure 4**

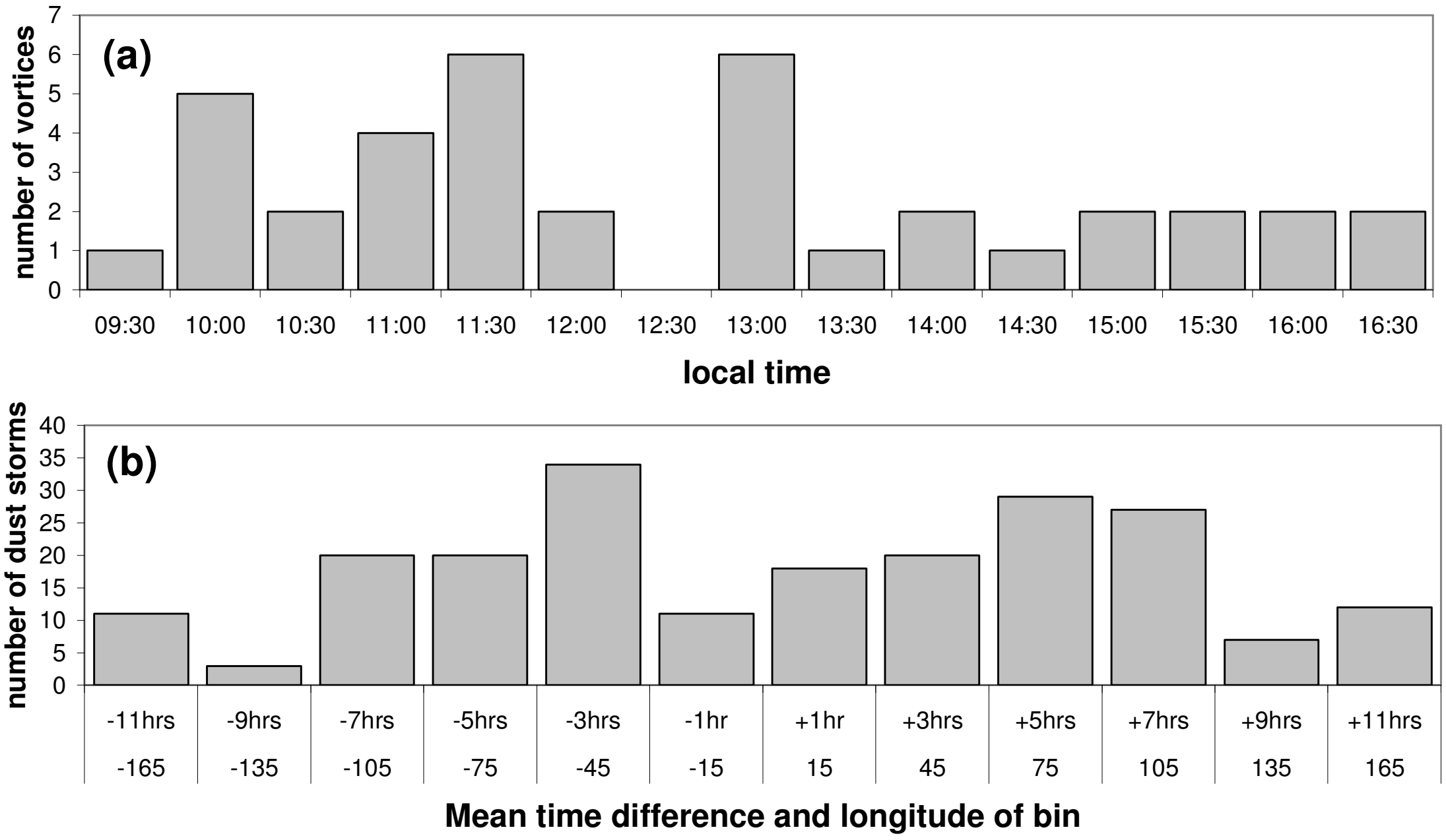

**Figure 5**

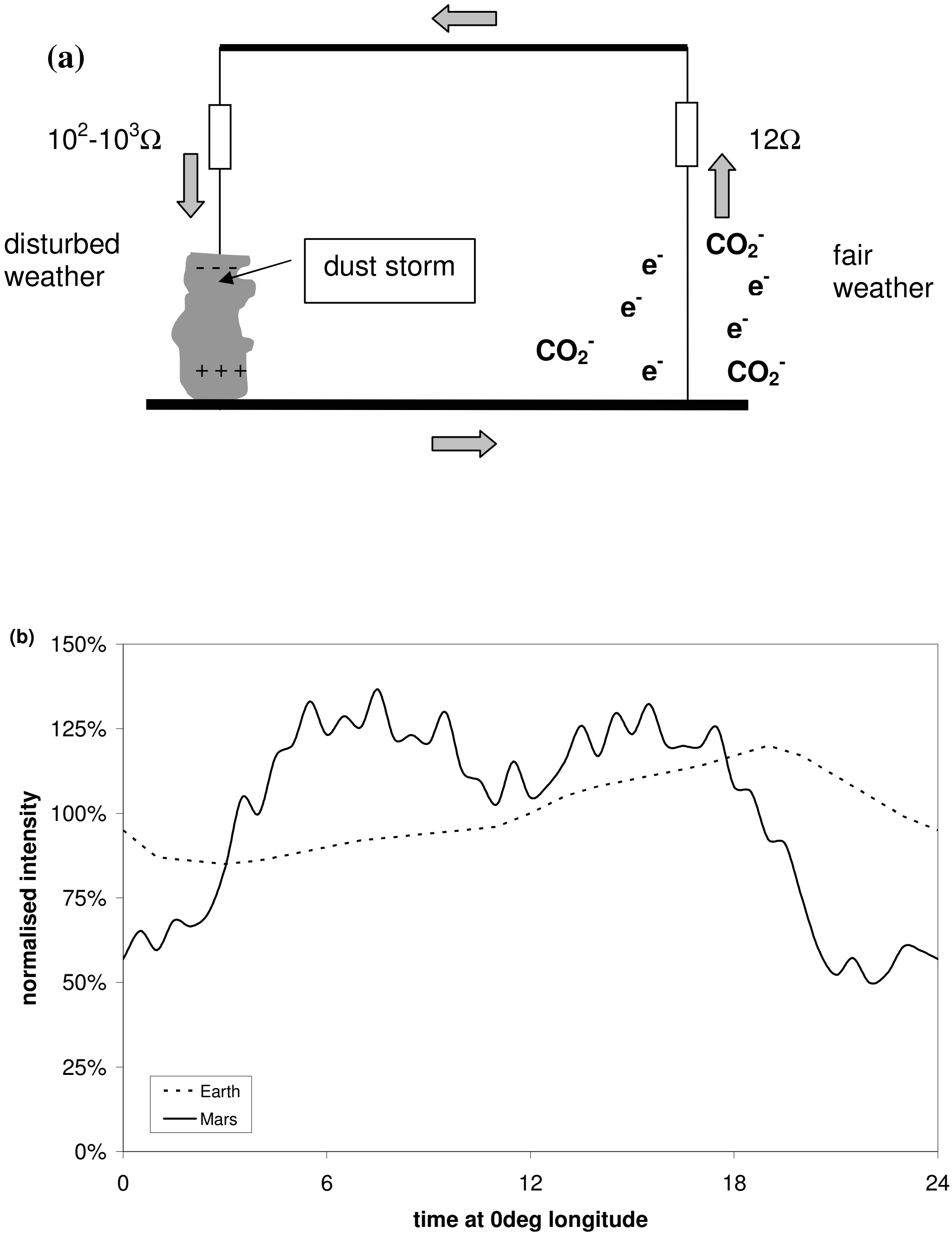

**Figure 6**

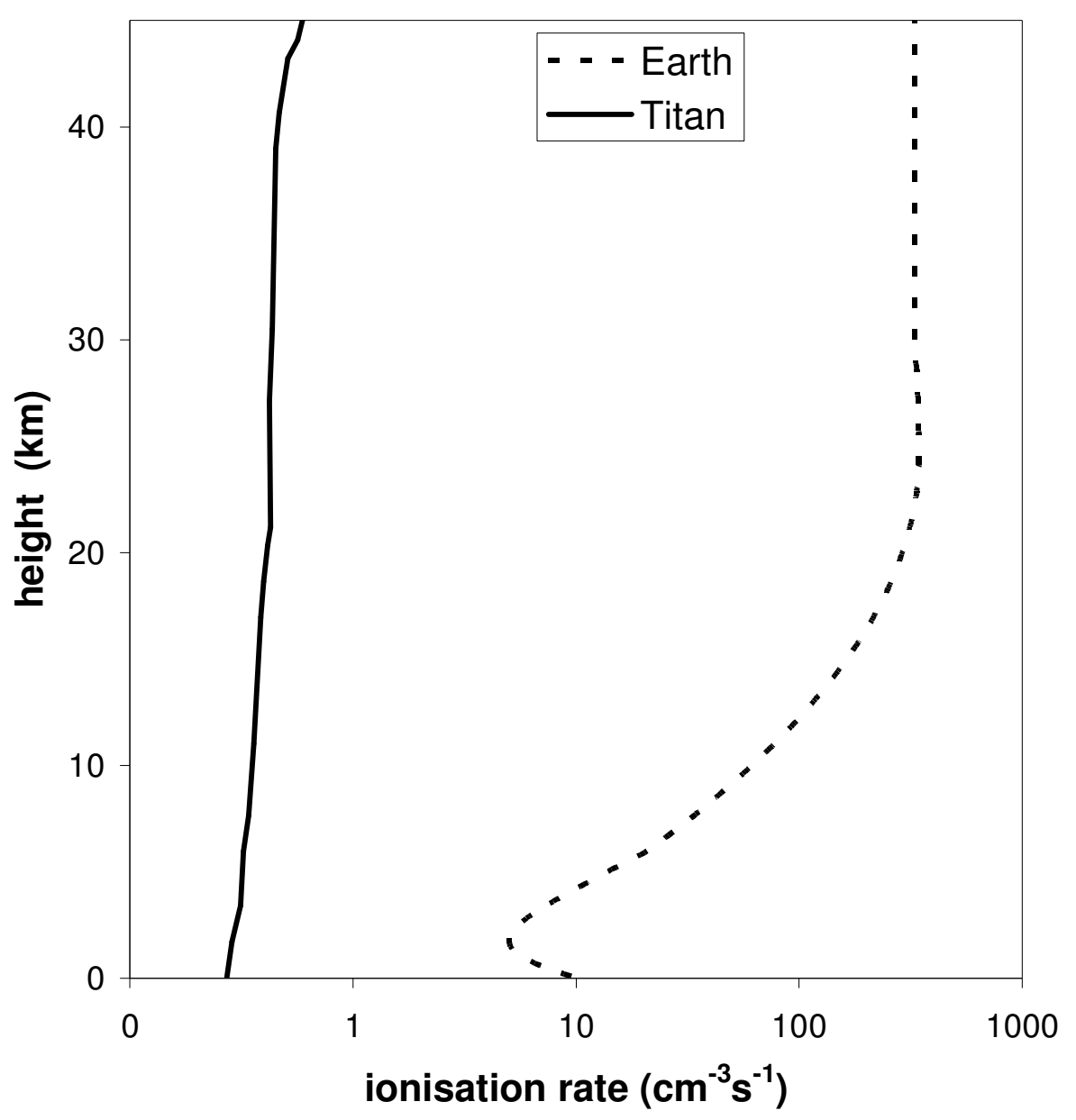

**Figure 7**